\documentclass{article}

\usepackage[preprint]{neurips_2025}

\usepackage[utf8]{inputenc} 
\usepackage[T1]{fontenc}    
\usepackage{hyperref}       
\usepackage{url}            
\usepackage{booktabs}       
\usepackage{amsfonts}       
\usepackage{nicefrac}       
\usepackage{microtype}      
\usepackage{xcolor}         
\usepackage{comment}
\usepackage{graphicx}
\usepackage{multirow}
\usepackage{paralist}
\usepackage{amsmath}
\usepackage{xcolor}
\newcommand{\rev}[1]{\textcolor{blue}{#1}}
\usepackage{float}
\usepackage{subcaption}
\usepackage{wrapfig}
\usepackage{listings}
\usepackage{pifont}

\title{POSITION: Collaboration Between the City and Machine Learning Community is Crucial to Efficient Autonomous Vehicles Routing}

\author{
    Anastasia Psarou$^{1}$,
    Ahmet Onur Akman$^{1}$,
    Łukasz Gorczyca$^{1}$,
    Michał Hoffmann$^{1}$,\\
    \textbf{Grzegorz Jamróz}$^{1}$,
    \textbf{Rafał Kucharski}$^{1}$\\
    \textnormal{$^1$ Faculty of Mathematics and Computer Science, Jagiellonian University, Kraków, Poland}\\
}

\begin{document}

\maketitle

\begin{abstract}

Autonomous vehicles (AVs), possibly using Multi-Agent Reinforcement Learning (MARL) for simultaneous route optimization, may destabilize traffic networks, with human drivers potentially experiencing longer travel times. We study this interaction by simulating human drivers and AVs. Our experiments with standard MARL algorithms reveal that, 
both in simplified and complex networks, policies often fail to converge to an optimal solution or require long training periods. This problem is amplified by the fact that we cannot rely entirely on simulated training, as there are no accurate models of human routing behavior. At the same time, real-world training in cities risks destabilizing urban traffic systems, increasing externalities, such as $CO_2$ emissions, and introducing non-stationarity as human drivers will adapt unpredictably to AV behaviors. 
\textbf{In this position paper, we argue that city authorities must collaborate with the ML community to monitor and critically evaluate the routing algorithms proposed by car companies toward fair and system-efficient routing algorithms and regulatory standards.}
\end{abstract}

\section{Introduction}

\begin{figure}[t]
\vspace{-20pt}
\begin{center}
\centerline{\includegraphics[width=\columnwidth]{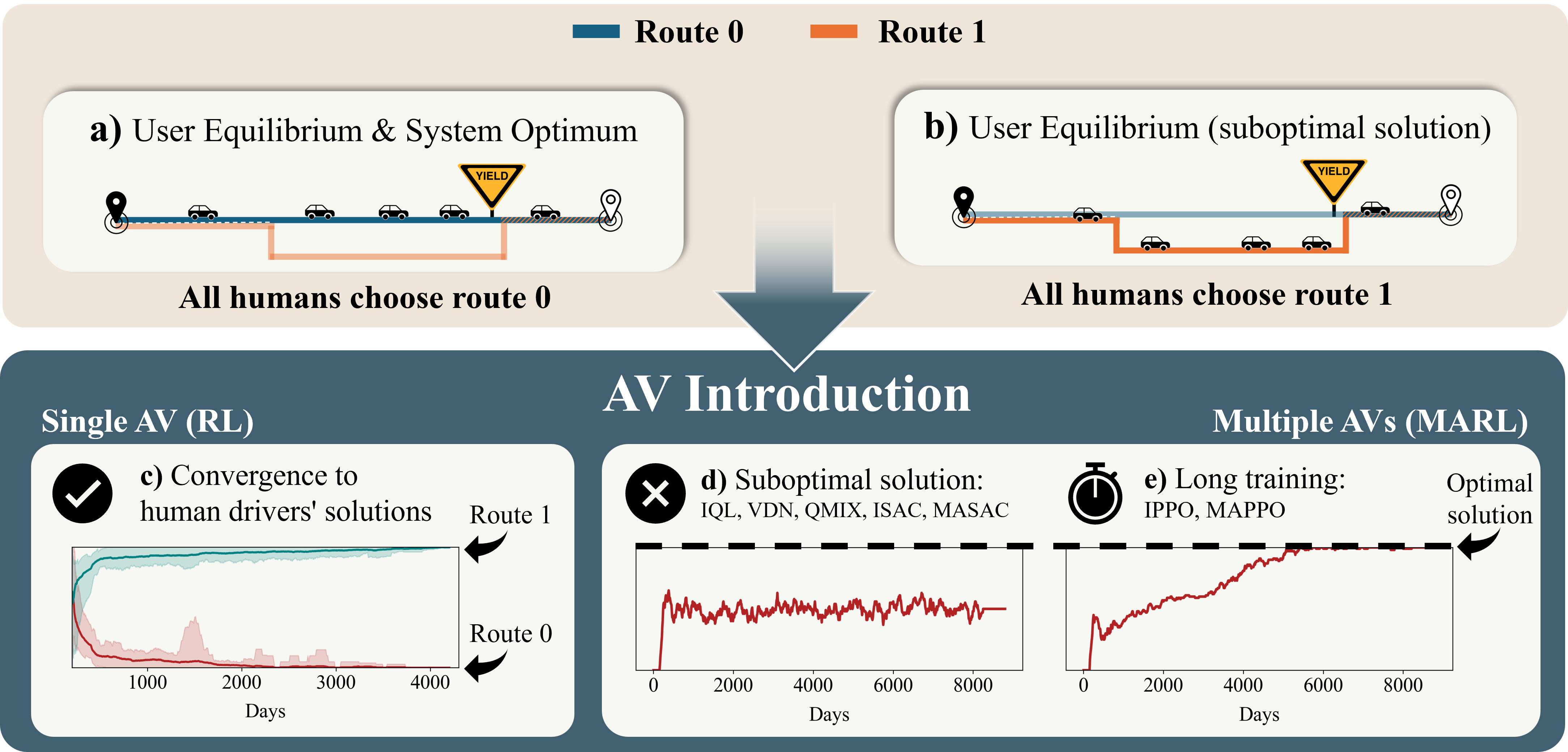}}
\caption{\textbf{Overview.} We demonstrate how MARL fails to find optimal routing policies on a simple topology of \textbf{T}wo \textbf{R}outes: shorter with no priority (\textbf{Y}ield sign) and longer with priority (\texttt{TRY} network). We simulate 22 human drivers who reach two equilibria: (a) optimal when all drivers use the shorter route 0, and (b) suboptimal when all use the longer route 1 (the first three humans always choose route 1 in this solution). If we replace a single human driver with an AV, any standard RL algorithm will quickly find the optimal routing policy (c). However, when multiple AVs (10) learn simultaneously, many \textbf{MARL algorithms fail to converge to the optimal policy} (d) or require hundreds of days (episodes) until they find it (e) in this trivial case as well as in real complex systems (Ingolstadt).}
\label{fig:overview}
\end{center}
\vspace{-2.2em}
\end{figure}

In urban traffic networks, humans (drivers) every day make routing decisions \cite{arnott1990departure} to arrive at their destinations as fast as possible \cite{bovy2005modelling}. With the advent of autonomous vehicles (AVs), these routing decisions may be delegated to algorithms, aiming to maximize the reward by selecting the optimal action in the current state of the network. Classically, this problem was formulated as a game-theoretical problem \cite{correa2004selfish}, where humans independently maximize their perceived payoffs.

As AVs will start sharing the roads with humans in a mixed system, at least for a while until human driving ceases, they will be influencing the complex social dynamics of individual, rational yet non-deterministic human driving behavior \cite{influence_of_AVs_on_car_following_behavior}. However, as we argue, current machine learning (ML) systems will not have (as of today) a sufficiently detailed model of urban mobility capable of training routing algorithms. Furthermore, as we demonstrate, specifically during training, the joint actions of AVs may lead to suboptimal solutions, resulting in a cost to all users in the capacitated system with limited resources. 
 Moreover, AVs are likely to take different actions than human drivers, which will likely trigger adaptations of humans that will change their routing policies in response to AVs' actions. This impact is negligible to some level, and single or few AVs routing recklessly during training will not disequilibrate the system. Yet the critical mass of AVs can be quickly reached as the AVs become broadly available (as little as 15\% of AVs in our experiment disequilibrate the system). This aligns with projections indicating rapid growth in AV deployment, with the number of commercial AVs for ridesharing expected to reach a few million by 2030 \cite{goldman2024partiallyautonomous}.

\textbf{This position paper argues that city authorities must begin actively collaborating with the ML research community to monitor and evaluate AI-based routing algorithms for AVs deployed by car companies. 
In parallel, the ML community should focus on continuously improving existing algorithms to ensure they are robust, fair, and suitable for real-world deployment. 
Without regulation, autonomous collective routing could introduce chaos into urban traffic systems and exacerbate congestion.} We support this position by demonstrating that, in simplified and real-world networks, when multiple AVs simultaneously learn 
routing policies using MARL, they will either destabilize the road networks and fail to find optimal solutions or learn long enough to affect the system's performance. While MARL is not the only viable approach, it is the one we focus on in this study. Our argument is supported by experiments conducted on a toy network (Figure \ref{fig:overview}a, b) and complemented with experiments on the real-world Ingolstadt network (Figure \ref{fig:ingolstadt_network}), using a portfolio of MARL algorithms. We analyze and show that:
\begin{list}{$\bullet$}{\leftmargin=1em \itemsep=0pt}
\item Human drivers, by maximizing their payoffs, stabilize the system into two equilibria (Figure \ref{fig:overview}a, b). 
\item RL efficiently finds the optimal route for a single AV (Figure \ref{fig:single-agent}). However, when multiple human drivers are replaced by AVs, MARL either \textbf{fails to converge or needs lengthy training} to find the optimal solution (Figure \ref{fig:marl-proportions}(a, b)).
\item The simulators of urban mobility are not ready to serve as virtual environments to train MARL, and training in the real world will be at the cost of the system's performance (Section \ref{sec:traffic_models}).
\item Optimal equilibrium state can easily transition to suboptimal (with worse performance for each agent) when non-determinism (e.g., in the traffic simulation) and/or human adaptation is introduced as another source of non-stationarity (Figure \ref{fig:marl-proportions}c).
\item Centralization can, in some cases, speed up convergence 
to optimal policies, but at the cost of our privacy (Figure \ref{fig:adapt-centr} bottom).
\item In the real-world Ingolstadt network, multiple simultaneous routing AVs can destabilize the system (Figure \ref{fig:ingolstadt-ippo}).

\end{list}

These phenomena can hinder the massive potential of AVs to contribute to sustainability \cite{taiebat2018review}, efficiency \cite{talebpour2016influence}, and optimality \cite{zhou2024modeling} of urban mobility. While Connected and Autonomous Vehicles (CAVs) 
promise novel routing strategies that allow the reduction of total and individual costs (travel times) and system costs (like total delays) \cite{jamróz2024socialimpactcavs} and its externalities ($\text{CO}_\text{2}$, $\text{NO}_\text{x}$, safety, noise, etc.) \cite{kopelias2020connected} - these benefits can be limited by the challenges described above. 

Based on experiment findings, to safely and reliably exploit the opportunities that AVs and MARL offer to the future urban traffic systems, \textbf{we call for}:

\begin{list}{$\bullet$}{\leftmargin=1em \itemsep=0pt}
    \item \textbf{The introduction of a regulatory framework}, developed in collaboration with the ML community, that requires car companies to submit their routing algorithms for certification before deployment in public road networks.
    \item \textbf{The development and deployment of monitoring systems} that track collective routing behaviors and can allow authorities to detect issues, like inequitable travel time distributions or prolonged system-wide congestion.    
    \item \textbf{The data-driven development of urban traffic simulators}, led jointly by ML and transportation research, to realistically reproduce human route choice (in the presence of real-time information) and its adaptation to dynamic environmental changes.
    \item \textbf{The continuous development and improvement of ML algorithms} that can robustly handle non-stationarity, scale with agent populations, and adapt to real-world traffic conditions.
    \item \textbf{Broad experimental studies} to benchmark routing algorithms across diverse traffic scenarios and identify failure modes that inform regulatory policies and guide safe algorithm design.
    
\end{list}

\section{Background}
\subsection{Partially Observable Markov Game (POMG)}
We abstract the daily (repeated) route choices made by humans (and AVs in the future) in capacitated networks (limited resources) to the so-called routing game. This, in turn, can be formalized with a POMG as a MARL problem. We denote a finite-horizon multi-agent general-sum POMG as a tuple $(\mathcal{S}, \{\mathcal{O}_i\}_{i \in \mathcal{N}}, \{\mathcal{A}_i\}_{i \in \mathcal{N}}, \mathcal{P}, \{r_i\}_{i \in \mathcal{N}}, \gamma)$, where $\mathcal{N} = \{ 0, \ldots, N-1 \}$ represents the set of $N$ agents. $\mathcal{S}$ represents a finite state space, partially observable by the agents. The global state $s \in  \mathcal{S}$ includes the joint agent observation (i.e., $o_{1} \times o_{2} \times ... \times o_{N}$) and additional environmental information. Each agent $i$ applies an action $a_{i}^t \in \mathcal{A}_i$ at each timestep $t$ from its own action space $\mathcal{A}_i \subseteq \mathcal{A}$ to the environment and consequently receives a reward $r_{i}^t$ from the environment as a function of the state and the joint action. The state transition probability is defined by $\mathcal{P}(s' \mid s, a): \mathcal{S} \times \mathcal{S} \times \mathcal{A} \to [0, 1]$ and indicates the probability of transitioning to state $s' \in \mathcal{S}$ after a joint action $a = (a_1, a_2, \dots, a_N) \in \mathcal{A}$.  A policy $\pi_i : \mathcal{O}_i \times \mathcal{A}_i \to [0, 1]$ specifies the probability distribution over the actions of agent $i$. The goal of each agent $i$ is to learn a policy, $\pi_i$, that maximizes its expected cumulative reward $G = \mathbb{E}_{\pi_i} \left[ \sum_{t=0}^{\infty} \gamma^t \cdot r_i^t \right]$, where $\gamma \in [0, 1)$ is the discount factor.

\subsection{Multi-agent reinforcement learning}
\label{sec:marl-motivation}

The route assignment 
is a combinatorial optimization problem. A plausible extension is to employ ML-equipped AVs and specifically use \emph{multi-agent reinforcement learning}, where each agent (vehicle) learns optimal policies to select the best route in the currently observed state of the road network. 
MARL involves multiple agents interacting within a shared environment, where each agent's actions can influence the state of the environment. This makes the environment non-stationary from a single agent's perspective. 
Using a single-agent RL algorithm to learn value functions of \textit{joint} actions would eliminate the non-stationarity \cite{centralization, ejal} but would not 
scale well when the size of the action space grows exponentially with the number of agents \cite{lu2024centralized}. 

One approach is to train a set of \textit{independent learners} (IL) where each learner treats other agents as part of the environment. 
We use Independent Q-Learning (IQL) as the initial baseline \cite{iql}, followed by two actor-critic methods, that is the Independent SAC (ISAC), the multi-agent version of the Soft Actor-Critic (SAC) algorithm \cite{sac}, and the Independent Proximal Policy Optimization (IPPO) algorithm, as it has demonstrated benchmark performance across a variety of problems \cite{yu2022surprisingeffectivenessppocooperative, papoudakis2021benchmarkingmultiagentdeepreinforcement}. 

Existing literature widely utilizes the \textit{Centralized Training and Decentralized Execution} (CTDE) structure, in which agents learn decentralized policies in a centralized manner, to be used independently during the execution phase \cite{lowe2020multiagentactorcriticmixedcooperativecompetitive}. This structure tackles partial observability while maintaining feasible computation times. Following this structure, \cite{vdn} introduced Value Decomposition Networks (VDN), a value-based algorithm that decomposes complex learning problems into sub-problems. \cite{qmix} proposed a similar algorithm named QMIX, which 
adopts a mixing network with a monotonicity constraint, enabling more complex coordination. Alternative to value decomposition methods, \cite{mappo} introduced a multi-agent version of the Proximal Policy Optimization (PPO) algorithm \cite{ppo}, Multi-Agent PPO (MAPPO). Similarly, Multi-Agent SAC (MASAC) is the multi-agent version of the SAC algorithm.  All these algorithms were used to demonstrate our position in the experiments.




\subsection{Limitations: Are traffic models ready to train RL algorithms?}
\label{sec:traffic_models}
We illustrate our position with a numerical simulation designed to replicate the real world and its complexity. However, the actual conditions in which AVs will be deployed are much more complex.
Namely, at the level of: 
\begin{list}{$\bullet$}{\leftmargin=1em \itemsep=0pt}
\item \textbf{Traffic flow}. We use Simulation of Urban MObility (SUMO) \cite{SUMO2018}, which applies Intelligent Driver Model (IDM) \cite{Treiber_2000} accurate, but not perfect microscopic model of traffic. Real traffic is less predictable and noisy and admits rare events like accidents \cite{chen2018exploring}.
\item \textbf{Demand patterns}. In this paper (except from Section \ref{sec:non-determinism}), we assume a fixed commute pattern every day. 
Namely, each agent has a fixed origin from which it departs every day at the same time to a fixed destination. However, in real systems, humans change departure times, work irregularly, and occasionally stay at home or travel to different destinations \cite{Gonz_lez_2008, horni2016matsim, Bhat1999}. 
\item \textbf{Route choices}. Humans are non-deterministic decision makers, and their choices are probabilistic with a significant impact of variation and heterogeneity on the choice probabilities (like in the Logit model \cite{ben1999discrete}).
\item \textbf{Action space}. In real networks, the number of feasible routes explodes quickly, reaching $10^{40}$ in many real-world examples \cite{frejinger2009sampling}. 
\end{list}

Unfortunately, all of the above are only roughly approximated with state-of-the-art transport system models. 
Although the setup discussed in Figure \ref{fig:overview} is simple, it already gives rise to issues (like increased travel time for humans), and these 
worsen as the system scales to bigger networks (replicated using the Ingolstadt network) with irregular demand patterns and non-deterministic traffic flows (see Figure \ref{fig:marl-proportions}c). 
In these cases, we get similarly bad results: MARL fails, supporting our position.

\section{Related work}
\label{sec:related_work}

\subsection{Analytical solutions} 

While the nature of the problem and its formalization as a POMG renders (MA)RL likely to be the tool of choice to solve fleet route choice problems we do not argue that it is the best and only solution. 
The classic traffic assignment solvers 
identify user equilibrium and system optimum using Operations Research (OR) methods, like Frank-Wolfe \citep{fukushima1984modified}. 
However, these methods operate on macroscopic flows 
and convex, continuous travel-time functions \cite{kucharski2017estimating}, 
while we study an
agent-based setting 
with a realistic, microscopic traffic flow model (SUMO). Alternatively, game-theoretical equilibria solvers, like Gambit \cite{turocy2001gambit}, can be used. 
However, Gambit requires a precomputed payoff matrix, that includes $2^{10}$ episode runs to account for all the possible joint action combinations 
and the complexity of the game grows exponentially. Hence, 
such solvers often struggle to adapt to the complex and dynamic nature of mixed traffic scenarios involving AVs and human drivers \cite{BAMDADMEHRABANI2024} and 
scale poorly in bigger networks, highlighting the need for alternatives, with MARL emerging as the most promising.

\subsection{Multi-agent reinforcement learning} 

 
RL has been applied to the route choice problem in a macroscopic setting, where vehicle groups are modeled collectively as flows 
instead of discrete agents. For example, \cite{Thomasini+2023} studied route choice in a centralized multi-agent setting using a macroscopic traffic simulator. Additionally, \cite{ZHOU2020124895} formulated route choice as a congestion game transforming it into the Traffic Assignment Problem (TAP). They proposed an RL-based solution that converges almost surely to the optimal solution aiming to minimize the total travel time in traffic networks. While macroscopic approaches 
explore overall traffic dynamics, they abstract away the heterogeneity and strategic behavior of individual agents.

In contrast, microscopic approaches focus on individual vehicle decisions, making them
more suitable for modeling AV–human interactions. For instance, \cite{AgentRewardShaping} adopted a 
multi-agent approach where drivers use Q-learning for route selection using reward-shaping techniques to reduce traffic congestion. Additionally, \cite{regret_route_choice} proposed a regret-minimization approach that relies on external traffic data. 

Some studies model route choice as a sequence of decisions made at each network node \cite{indi_vs_diff_rewards_route_choice, improved_learning_automata_route_choice}, allowing for dynamic adaptation to changing traffic conditions. In a different vein, \cite{lazar2021learningdynamicallyrouteautonomous} consider a setting where human drivers act selfishly and AVs centrally controlled using deep RL decrease congestion by indirectly influencing human's routing decisions. Lastly, \cite{akman2024impact} introduced AV-specific behavioral reward formulations in mixed-traffic environments which are later included in the RouteRL framework \cite{RouteRL2024} for
simulating the collective route choices of both human drivers and AVs.

\section{Problem statement}
\label{sec:components_of_simulation}

\begin{wrapfigure}{r}{0.5\textwidth}
    \vspace{-1.5em} 
    \centering
    \includegraphics[width=\linewidth]{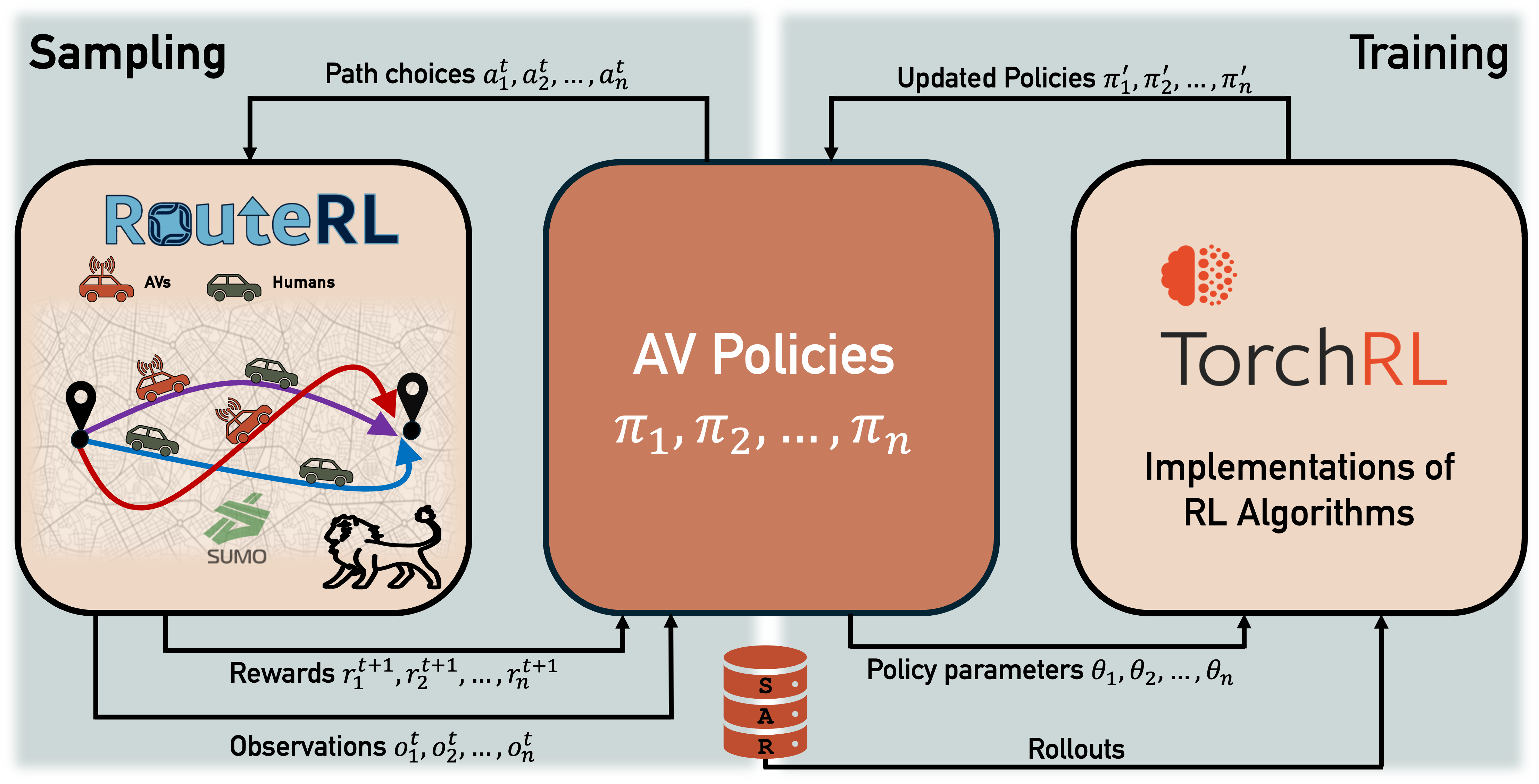}
    \caption{We model the routing "game" between humans and AVs on urban traffic networks using the \textbf{RouteRL} \cite{RouteRL2024} framework, which includes a custom \textit{PettingZoo} \cite{terry2021pettingzoo} environment, 
    communicating with \textit{SUMO} as it trains optimal routing policies with standard MARL algorithms via \textit{torchRL} \cite{bou2023torchrl}.
}
    \label{fig:routerl}
    \vspace{-0.7em} 
\end{wrapfigure}

We illustrate our position on a simple \textbf{T}wo-\textbf{R}oute (\textbf{Y}ield) network (\texttt{TRY}) (Figure \ref{fig:overview}a, b), which, while 
abstract, 
is carefully designed to capture 
arising issues of AV introduction in traffic networks. 
To assess the generalizability of our findings we also include a real-world case. In the \texttt{TRY} network, each agent selects from two possible routes (0 or 1) to reach the destination. Then, the reward (travel time) is collected from the environment to update choices for the next episode. Each episode is interpreted as a day on which the 22 drivers commute through the network. For clarity, the setting is unrealistically static. Each human driver follows the same route every day, departing at the same time, and the travel times do not vary day-to-day (none of the above holds in real networks as discussed in \ref{sec:traffic_models}, see Appendix \ref{sec:SUMO_appendix}). By design, the system has two Nash equilibria achievable by humans, one optimal (System Optimal - User Equilibrium, Figure \ref{fig:overview}a) and one suboptimal (Suboptimal - User equilibrium, Figure \ref{fig:overview}b). Humans \emph{mutate} into AVs, which will use any MARL algorithms to find optimal policies. To simulate this scenario, we use the RouteRL \cite{RouteRL2024} framework (Figure \ref{fig:routerl}), which models mixed-traffic scenarios, where AVs are simulated as RL agents and humans behave according to a given human-behavior model.

\textbf{SUMO.} An open-source, state-of-the-art, microscopic, agent-based traffic simulator used as the traffic environment in which each vehicle navigates the road network according to the IDM \cite{SUMO2018}. 

\textbf{Action.} The action space of the agents corresponds to the set of available routes connecting their given origin with their destination and is discrete with value two on the TRY network (See \ref{fig:overview}a, b).

\textbf{Reward.}
The reward $r_{i}^t(a_{i}^t | s_{i}^t)$ is the negative travel time of each agent $i$ to reach from its origin to its destination, as calculated by SUMO.

\textbf{Observation.}
We assume, plausibly for the future systems, that AVs observation \(o_{i}^t\) is composed of their departure time and the number of agents that departed before them and chose routes 0 and 1.

\textbf{Human agents.}
\label{par:human_agents}
We follow the classical representation of human route-choice behavior from transportation research. Human drivers are rational decision-makers aiming to maximize their perceived utility \cite{Cascetta} by selecting actions that minimize expected travel times. Their expectations are updated based on experienced travel times (from SUMO). In scenarios with adaptation, humans shift to an alternative route with $10\%$ probability. 

\subsection{Human system and its equilibria.}
\label{sec:equilibria}

We consider two plausible equilibria resulting from human collective actions: first, when all humans select the shorter route ($[\{0\}^{22}]$, \ref{fig:overview}a), and second when they all opt for the longer route ($[\{1\}^{22}]$, \ref{fig:overview}b).
Both meet Nash criteria for User Equilibrium \cite{Wardrop1952ROADPS} (common paraphrase of Nash equilibrium for the route choice context), with the former being also System Optimal (minimizing total travel time in the system \cite{merchant1978optimality}). None of the drivers is inclined to change their route, as it would reduce individual rewards (travel longer and arrive later). For stability, we fix the route for the first three agents to regulate the early loading of the system. After 200 days of simulation, the system is stable (the next day the rational drivers will replicate today's choices), fair (travel times are equal among drivers), and either globally optimal (Figure \ref{fig:overview}a where total travel costs are minimized) or suboptimal (Figure \ref{fig:overview}b). The individual and system costs (travel times) are reported in Table \ref{tab:costs}.

\section{Experimental support for the position}

\subsection{Single AV routing with RL.}
\label{sec:single_agents}

In the equilibrated human system, we first replace a random human vehicle with an RL-controlled AV. Its traffic properties (reaction, acceptance gap, acceleration profile, and other IDM  parameters \cite{Treiber_2000}) remain intact. AVs are indistinguishable from humans by all but routing decisions: they may use any RL (MARL) algorithm to converge to the optimal policy.

\begin{wrapfigure}{r}{0.35\textwidth}
    \vspace{-2em} 
    \centering
    \includegraphics[width=\linewidth]{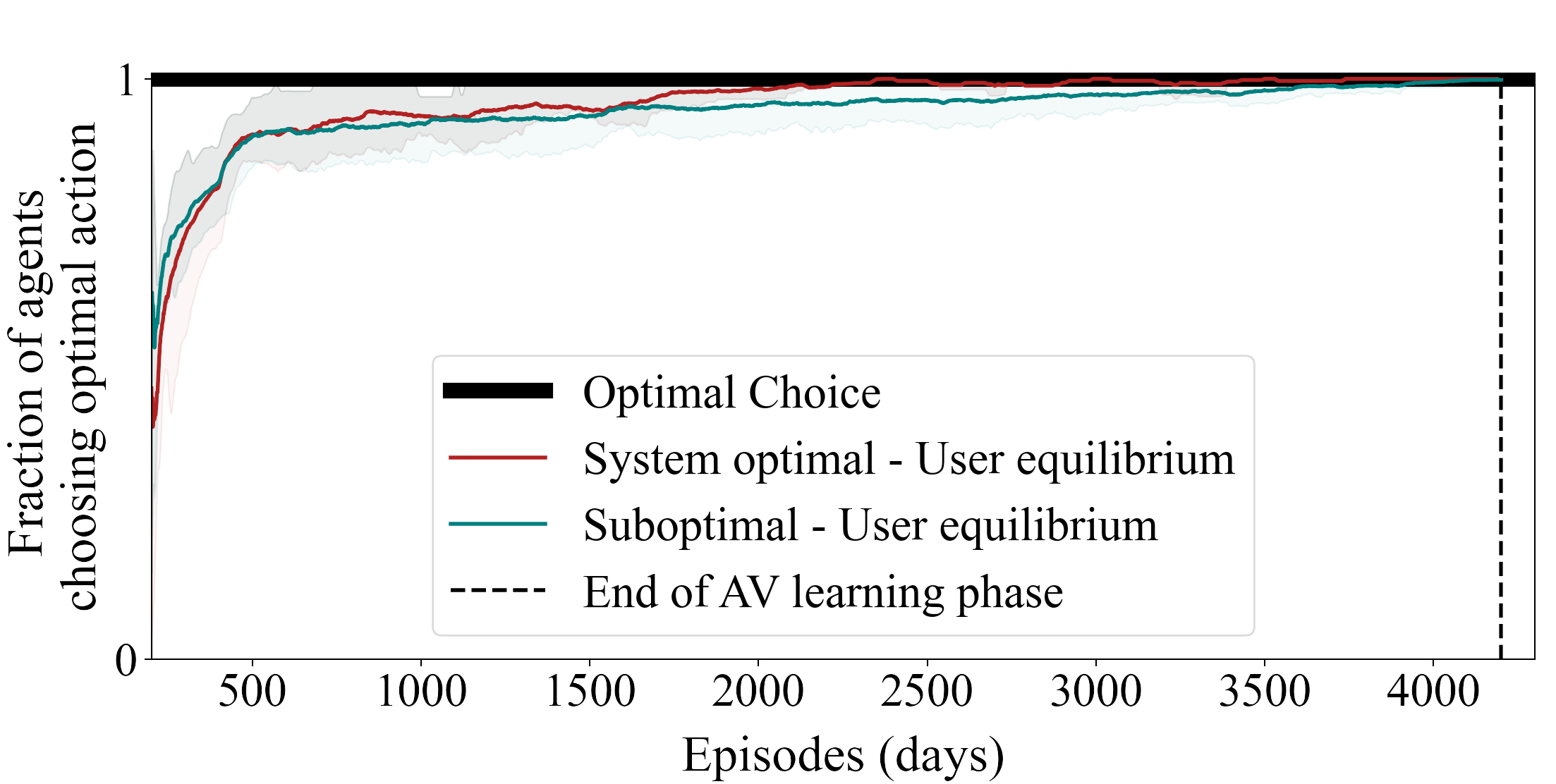}
    \vspace{-0.5em} 
    \caption{\textbf{Single AV} replicates human choices, and each tested RL algorithm finds the optimal solution to the binary choice problem (means and error bars computed across replications; see Section \ref{sec:plot_description}).}
    \label{fig:single-agent}
    \vspace{-1em} 
\end{wrapfigure}

We demonstrate that RL 
finds the optimal solution, and AV agents follow the crowd (replicate the decision of the human that drove that vehicle before). Unsurprisingly, since the problem is a trivial binary decision in a static environment, this is true for all suitable RL algorithms, including Deep Q-Learning (DQN), PPO, and SAC, as we show in Figure \ref{fig:single-agent}. In any equilibrated system, any RL algorithm training AVs with the same reward formulation, and action space as their human predecessors will replicate the optimal strategy, which can be derived directly from conditions of Nash equilibrium. Moreover, in a dense traffic environment, the impact of a single vehicle is unlikely to be noticed by other humans. The marginal cost of actions taken by a single AV to other humans will fall within what is known as the indifference band \cite{di2017indifference}, making them indistinguishable from the traffic stochastic noise \cite{neun2023traffic4cast}. \textbf{Our position, that AVs will disequilibrate the traffic networks, does not hold for a single AV}, that will converge to the solution of its human predecessor and will not impact other human drivers.

\subsection{Simultaneous learning of multiple AVs with MARL.}
\label{sec:multi-avs}
Already in our simple scenario, serious issues arise when multiple vehicles learn optimal policies simultaneously. Most likely, more than one AV will be introduced into our cities (whenever they become ready to operate reliably), each solving the same problem of identifying optimal routing policies to reach its destination. Such a multi-agent setting allows a significant alteration to the initial problem: the AVs may communicate (becoming the so-called CAVs) and share information.

First, we demonstrate the natural first stage of AV introduction: with no communication and reward formulation identical to the one of the selfish human drivers. Now, the environment has become non-stationary \cite{jiang2024blackbox}, as the state transitions and rewards of an agent are influenced by the evolving policies of other agents. With as few as 10 AVs (where non-convergence appears after the introduction of just 4 AV agents discussed in Section \ref{para:fleet_size} and Appendix \ref{sec:critical_fleet_analyxix_appendix}, \ref{sec:smaller_av_fleets}) our setting is sufficient to argue for our \emph{position}. 

\textbf{Despite the very small size of joint action space, some algorithms fail to find the optimal solutions after thousands of iterations. Others need hundreds of policy updates to find the trivial solution}. Specifically, the trivial solutions are $[\{1\}^{22}]$ or $[\{0\}^{22}]$ depending on the human equilibrium from which we start, and the joint action space is $2^{10}$.

\begin{figure}[ht]
\centering
\includegraphics[width=1\linewidth]{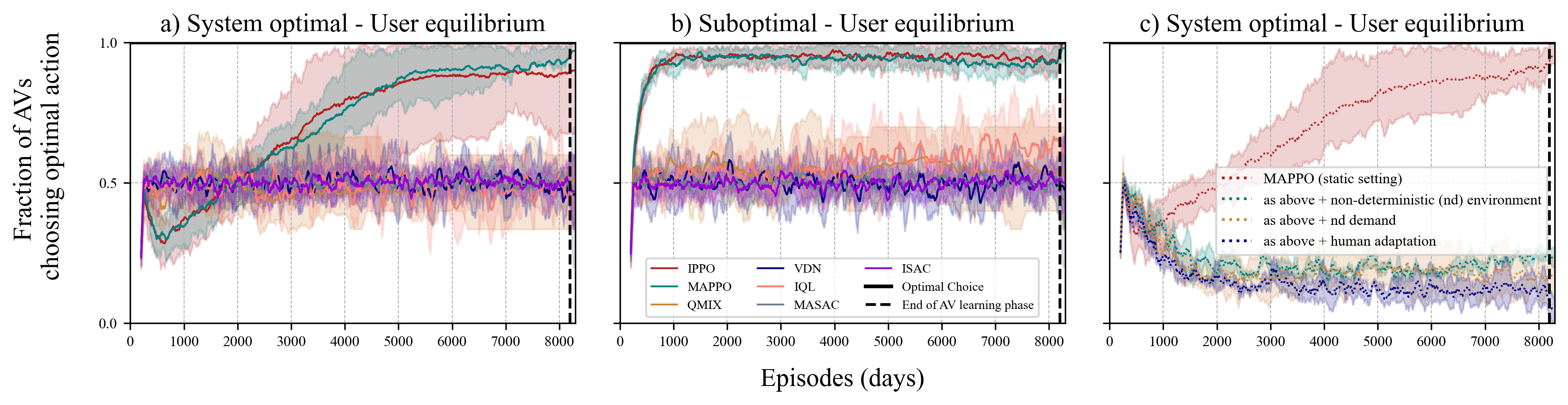}
\caption{\textbf{(a, b): 8000} episodes (that correspond to over \textbf{20 years}) 
of training for selected MARL algorithms. The trivial solution is found only by IPPO and MAPPO. Other algorithms fail to converge to the optimal policy. Convergence in the suboptimal equilibrium is achieved after 1000 days, but System optimal requires much more training, far exceeding the patience threshold for the remaining 12 human drivers in the system. 
The first 200 episodes represent the human learning phase, followed by the training and testing phases of the AVs. \textbf{(c):} The MAPPO solution (red), however, is not robust and falls to suboptimal as soon as non-determinism is introduced to the environment, demand, or human behavior (other lines).
The line represents the mean with 0.15 and 0.85 percentiles. See Appendix  \ref{sec:experiments_details} for the experiments' setup.}
\label{fig:marl-proportions}
\vspace{-18pt}  
\end{figure}

In Figure \ref{fig:marl-proportions}(a, b), we present two arguments supporting our position. First is the class of algorithms that failed to converge after sufficiently long training (QMIX, VDN, IQL, MASAC, ISAC). Specifically, the choices are far from optimal, as reflected in the noisy profiles that fluctuate around 0.5. QMIX and VDN probably do not find optimal solutions due to the cooperative nature of their implementation. The difficulty of achieving stability and convergence in ISAC and MASAC can be attributed to their off-policy nature combined with the nonlinear function approximation used in neural networks \cite{DBLP:journals/corr/abs-1812-05905, NIPS2009_3a15c7d0}. The consequences of introducing MARL routing algorithms to AVs are negative to all parties: not only the travel times are longer for all agents (humans and AVs) but they also become variable, as we report in Table \ref{tab:costs}. IPPO and MAPPO, however, converged after a lengthy training. Theoretical explanations of the convergence difficulties are provided in Appendix \ref{sec:theoretical_analysis}.

\subsection{Training in virtual traffic environment.} Some MARL algorithms converged and successfully managed to identify optimal policies (the small joint action space of 1024 can be easily enumerated to explicitly identify optimal solutions). Eventually, the 10 AVs managed to stabilize their actions and reliably make optimal choices (Figure \ref{fig:marl-proportions}(a,b)). After that, the negative impact on the system and the humans diminishes (see Table \ref{tab:costs}). Nonetheless, the learning process was lengthy, with thousands of episodes (days).

The training can be interpreted twofold: within a virtual environment or in a real system. The former is neutral to the real system and its users since AVs can train their policies virtually and deploy them only after training is complete and the policies have converged to optimal solutions. Humans will not be affected and iterations remain only virtual. This, however, requires \emph{virtual environment} suitable to train a policy, which, is not available to date, as we argued in Section \ref{sec:traffic_models}. 
If so, the \textbf{learning period needs to be treated physically} and algorithmic iterations are not abstract episodes anymore, but physical days of real disturbances. Eventually, disturbances diminish (as for IPPO and MAPPO), yet the negative impact on the system and its users accumulates (to values presented in Section \ref{sec:cumulative_time_differences}). Alternatively, RL can be inaccurately trained on imperfect models and collected data and later fine-tuned in reality, as demonstrated in \cite{nair2021awacacceleratingonlinereinforcement}. However, \textit{sim2real} transfer presents significant challenges, as discussed in \cite{zhao2020sim}. Real-world complexity exceeds any simulation’s capabilities.

\begin{table*}[t]
\caption{Average travel times (rewards) (in seconds) for each subgroup (AVs, humans, and both), with standard deviations within each subgroup in parentheses. ‘Human system’ refers to rollouts up to the 200th episode, before the introduction of AVs. The remaining values (MARL, Centralized) are calculated from aggregated results during the testing phase and averaged across repeated experiment folds. The lowest travel times for each subgroup in each experimental setting are highlighted in \textbf{bold}.}
\label{tab:costs}
\begin{center}
\begin{sc}
\resizebox{\textwidth}{!}{
    \begin{tabular}{@{}llcccccc@{}}
    \toprule
    \multicolumn{1}{c}{} & \multicolumn{1}{c}{} & \multicolumn{3}{c}{System Optimum \& User Equilibrium} & \multicolumn{3}{c}{Suboptimal \& User Equilibrium} \\ 
    \multicolumn{1}{c}{} & \multicolumn{1}{c}{} & AVs & Humans & \multicolumn{1}{c}{All} & AVs & Humans & All \\ 
    \midrule
    {Human system} & \multicolumn{1}{c|}{} & - & 53.1 (13.1) & \multicolumn{1}{c|}{53.1 (13.1)} & - & 65.9 (15.5) & 65.9 (15.5) \\ 
    \midrule
    \multirow{7}{*}{MARL}
     & \multicolumn{1}{l|}{IPPO} & 59.1 (13.3) & 55.9 (21.1) & \multicolumn{1}{c|}{57.4 (18.2)} & \textbf{69.9 (13.3)} & \textbf{62.5 (16.4)} & \textbf{65.9 (15.5)} \\
     & \multicolumn{1}{l|}{MAPPO} & \textbf{57.4 (12.0)} & \textbf{51.2 (15.6)} & \multicolumn{1}{c|}{\textbf{54.0 (14.4)}} & \textbf{69.9 (13.3)} & 62.6 (16.4) & \textbf{65.9 (15.5)} \\
     & \multicolumn{1}{l|}{ISAC} & 72.1 (17.4) & 71.5 (26.9) & \multicolumn{1}{c|}{71.8 (23.1)} & 82.0 (16.7) & 61.0 (15.4) & 70.6 (19.5) \\
     & \multicolumn{1}{l|}{MASAC} & 69.3 (15.3) & 70.1 (24.7) & \multicolumn{1}{c|}{69.7 (21.1)} & 84.2 (18.0) & 60.7 (15.5) & 71.3 (20.4) \\
     & \multicolumn{1}{l|}{QMIX} & 67.9 (14.8) & 66.2 (24.1) & \multicolumn{1}{c|}{67.0 (20.5)} & 77.5 (17.8) & 60.9 (15.2) & 68.5 (18.5) \\
     & \multicolumn{1}{l|}{VDN} & 69.1 (16.3) & 67.1 (27.4) & \multicolumn{1}{c|}{68.0 (23.0)} & 82.2 (17.1) & 60.2 (14.6) & 70.2 (19.4) \\
     & \multicolumn{1}{l|}{IQL} & 68.7 (16.4) & 67.9 (25.0) & \multicolumn{1}{c|}{68.3 (21.8)} & 80.0 (16.5) & 61.1 (15.2) & 69.7 (18.5) \\ 
     \midrule
    \multirow{7}{*}{\begin{tabular}[c]{@{}l@{}}MARL\\ (Adaptation)\end{tabular}}
     & \multicolumn{1}{l|}{IPPO} & \textbf{65.3 (12.6)} & 74.1 (30.4) & \multicolumn{1}{c|}{70.1 (24.4)} & \textbf{69.7 (13.4)} & 65.4 (18.9) & 67.4 (16.8) \\
     & \multicolumn{1}{l|}{MAPPO} & 65.6 (12.2) & 78.0 (29.9) & \multicolumn{1}{c|}{72.3 (24.4)} & 69.8 (13.4) & 65.5 (18.9) & \textbf{67.4 (16.7)} \\
     & \multicolumn{1}{l|}{ISAC} & 70.2 (16.8) & 66.4 (23.9) & \multicolumn{1}{c|}{68.1 (21.1)} & 79.4 (17.9) & 63.5 (17.8) & 70.7 (19.7) \\
     & \multicolumn{1}{l|}{MASAC} & 71.6 (17.4) & 68.8 (26.8) & \multicolumn{1}{c|}{70.0 (23.1)} & 84.4 (17.9) & 62.9 (17.8) & 72.7 (20.9) \\
     & \multicolumn{1}{l|}{QMIX} & 68.5 (16.2) & \textbf{63.2 (23.2)} & \multicolumn{1}{c|}{\textbf{65.6 (20.5)}} & 80.6 (18.8) & 63.2 (18.1) & 71.1 (20.4) \\
     & \multicolumn{1}{l|}{VDN} & 72.2 (17.6) & 72.1 (26.0) & \multicolumn{1}{c|}{72.2 (22.6)} & 83.9 (17.3) & \textbf{62.7 (17.8)} & 72.4 (20.7) \\
     & \multicolumn{1}{l|}{IQL} & 68.6 (16.8) & 63.6 (24.2) & \multicolumn{1}{c|}{65.9 (21.3)} & 82.4 (17.0) & 63.4 (17.9) & 72.0 (20.2) \\ 
     \midrule
    \multirow{2}{*}{Centralized}
     & \multicolumn{1}{l|}{IPPO} & \textbf{64.7 (10.6)} & \textbf{83.9 (30.3)} & \multicolumn{1}{c|}{\textbf{75.2 (25.3)}} & \textbf{41.9 (8.0)} & \textbf{37.5 (9.8)} & \textbf{39.5 (9.3)} \\
     & \multicolumn{1}{l|}{MAPPO} & \textbf{64.7 (10.6)} & \textbf{83.9 (30.3)} & \multicolumn{1}{c|}{\textbf{75.2 (25.3)}} & 69.9 (13.3) & 62.5 (16.4) & 65.9 (15.5) \\ 
    \bottomrule
    \end{tabular}
}
\end{sc}
\vspace{-16pt}
\end{center}
\end{table*}

\subsection{Critical fleet size analysis.}\label{para:fleet_size}

\begin{wrapfigure}{r}{0.4\textwidth}
    \vspace{-3em} 
    \centering
    \includegraphics[width=\linewidth]{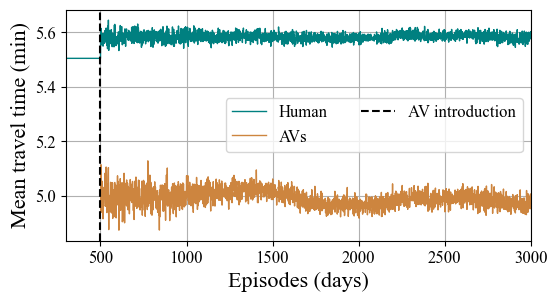}
    \vspace{-0.5em} 
    \caption{Introducing AVs on the \textbf{Ingolstadt} network (with 1000 agents and 1000 nodes) not only increased human travel times but introduced variability of travel times during training, which did not converge in \textbf{3000} days (MAPPO episodes).
    (see Appendix \ref{sec:ingolstadt_network}).}
    \label{fig:ingolstadt-ippo}
    \vspace{-1em} 
\end{wrapfigure}

As demonstrated, the problem behind our position is in multiple simultaneously learning agents. To find how many simultaneously learning agents can negatively affect the system, we simulated scenarios with a gradually increasing number of agents and reported when convergence issues arose. In each simulation, the specific AVs were chosen at random, with the condition that no two AVs are consecutive (there is always a human agent between AVs), and the first two vehicles never mutate. The algorithms start failing to converge even with 3-5 agents, as shown in Figure \ref{fig:different-fleet-number} and Appendix \ref{sec:smaller_av_fleets}. 
\textbf{This reveals a systemic issue that becomes increasingly prevalent as the number of agents grows, which will pose a serious threat to traffic performance when more AVs participate in our daily commute.}

\subsection{Privacy of personal data in centralized systems.}
\label{sec:centr}

Another aspect of our position lies in the communication, collaboration, and/or centralization \cite{schwarting2019social} - which makes AVs the \textbf{C}AVs. This allows us to better exploit the potential of autonomous driving, presumably at the cost of sharing private information with others or with central agents \cite{nayak2021autonomous}. Can we trade our private destination and origin to resolve the non-stationarity issues? To some extent, centralization may speed up the convergence, yet nowhere close to solving the problem and leading to issues with combinatorically growing search space (see Figure \ref{fig:adapt-centr} bottom in Appendix \ref{sec:centr-adapt-appendix}).


\subsection{Scaling from the abstract case to real-world scenarios - Ingolstadt results.}
\label{sec:non-determinism}

To reinforce our results, we reproduce our experiment and investigate what happens when complexity is added to the abstract case. We gradually introduce real-world phenomena (see Figure \ref{fig:marl-proportions}c and Table \ref{tab:determinism}) from Section \ref{sec:traffic_models}, successively including non-determinism in the traffic flow model (green), demand patterns (implemented as random departure times, yellow) and human adaptation (blue). 

As these complexities are introduced, the system becomes increasingly disequilibrated. In response,  humans will naturally seek ways to improve their payoffs (arrive faster) \cite{watling2013modelling}. This behavior is similar to the process of finding the equilibrium \cite{bie2010stability}, but even less predictable \cite{he2012modeling}. This adds another source of non-stationarity to the system. 
We include a probabilistic adaptation formula (to the human decisions), and 
the previously optimal system now shifts to the suboptimal state (Figure \ref{fig:marl-proportions}c).

\begin{wraptable}{r}{0.53\textwidth}
    \vspace{-1.3em} 
    \centering
    \caption{MARL convergence under varying complexity levels (\ding{51} = deterministic, \ding{55} = non-deterministic).}
    \vspace{-0.5em} 
    {\scriptsize
    \begin{tabular}{l|c|c|c|c}
        \toprule
        Network & Traffic Flow & Demand & Adaptation  & Convergence \\
        \midrule
        TRY & \ding{51} & \ding{51} & \ding{51} & yes \\
        TRY & \ding{55} & \ding{51} & \ding{51} & no \\
        TRY & \ding{55} & \ding{55} & \ding{51} & no \\
        TRY & \ding{55} & \ding{55} & \ding{55} & no \\
        Ingolstadt & \ding{51} & \ding{51} & \ding{51} & no \\
        \bottomrule
    \end{tabular}
    }
    \vspace{-0.5em} 
    \label{tab:determinism}
\end{wraptable}

Finally, we report the results of the Ingolstadt network, using realistic demand, from the RESCO benchmark \cite{ault2021reinforcement} (see Section \ref{sec:ingolstadt_network} for more details). This demand includes 1000 agents with varying origins and destinations, each selecting between four paths, rendering the joint action space to $4^{1000}$. As shown in Figure \ref{fig:ingolstadt-ippo}, the average travel time of all the human agents in the system (pre-AV) is stable. However, after the introduction of AVs (episode 500), the previously stable system becomes unstable, exhibiting oscillatory travel times with values higher than before.

In all the above-mentioned cases, the \textbf{central issue of our position remains present:} agents did not converge to the optimal solution even after many episodes (\ref{fig:marl-proportions}c), and during MARL training the system was destabilized (demonstrated as the variability of travel times on Figure \ref{fig:ingolstadt-ippo}).

\section{Conclusion}

AVs have begun appearing in traffic networks worldwide, and their presence is expected to increase in the coming years. As AVs become more prevalent, algorithms can be applied to optimize their route choices. 
In practice, the state-of-the-art MARL algorithms employed in this paper need hundreds of episodes to converge to optimal policies, even in trivial cases. The problem is amplified when more realistic traffic dynamics are introduced in the simulations. To realistically simulate human route-choice behavior, we miss theories, data to verify them, and simulation tools. Realistic urban mobility simulators are lacking, and the training episodes would probably need to be deployed directly on real traffic systems, disrupting the traffic networks. Needless to say, such a disruption should be avoided at all costs. 
We hope our contribution will spark discussions within the MARL community, encouraging the collaboration of authorities with the ML community to regulate autonomous collective routing.
\textbf{Such an experimental research program is needed 
to ensure we fully exploit the opportunities of the new technology (AVs) and algorithms 
to help us improve the traffic in future cities.}

\section{Alternative Views}
\label{sec:alternate_views}

This section covers some alternative perspectives that challenge our position.

\textbf{Why regulate AV routing since it is not yet a problem?}
Safety issues are indeed the foremost concern in AV deployment, and it may well be more than a decade before the problems indicated become urgent. Nevertheless, if this happens when large fleets are allowed to group routes, it may completely disrupt the traffic that should be preempted and now is the right time to act.

\textbf{Authorities should not monitor AI/ML-based routing algorithms.} 
One could argue that authorities do not need to monitor AI/ML-based routing algorithms, since cities already manage infrastructure through traffic signals, congestion pricing, and physical designs. Furthermore, the competitive market incentivizes companies to develop efficient routing tools to offer superior services to their users. While these points are valid, history shows that markets may degenerate into vicious degrading competition if left alone, especially when novel strategies, whose efficiency may involve dumping, introducing chaos, etc. become available to group players.


\textbf{What if car companies would not apply AI/ML for the routing policies of AVs?}
Artificial intelligence has already been integrated into many aspects of our daily lives. As discussed in Section \ref{sec:related_work}, reinforcement learning has been applied to optimize vehicle routing. It is therefore natural to expect that such technologies will also be adopted by car companies.

\textbf{What if reinforcement learning is not suitable for vehicle routing?} RL, as discussed in Section \ref{sec:related_work}, is already regarded as a viable solution for routing optimization, offering the ability to learn complex coordination and solve dynamic problems. However, this paper does not claim that MARL is the only method car companies might adopt. On the contrary, we argue that companies may deploy a range of algorithmic approaches—some of which could have even more detrimental effects on traffic flow. This is why such routing algorithms must be subject to monitoring.

\textbf{What if the limitations of the virtual environment are specific to this study and could be resolved with better design or tools?} As discussed in Section \ref{sec:traffic_models}, accurately representing dynamic, multi-agent systems like traffic networks requires highly complex simulation environments. Modeling realistic traffic demand patterns involves acquiring and handling data about real-world driver behavior, which is often private and sensitive.

\section*{Impact Statement}
\label{sec:impact}
\textbf{Our position aims to 
direct the research focus towards assessing the algorithms that can optimize the collective routing of AVs.} It demonstrates the potentially negative consequences for our future urban systems and their sustainability that are likely to occur if actions are not taken. We list gaps in the state-of-the-art that could render the large-scale deployment of AVs using MARL for urban routing decisions harmful to the overall system performance. To address these challenges, we propose research directions to leverage the capabilities of CAVs. We believe it is time to act now before the damage has been done.

\bibliographystyle{plain}  
\bibliography{references}

\begin{thebibliography}{10}

\bibitem{Janux}
Ahmet~Onur Akman.
\newblock Janux, 2025.

\bibitem{akman2024impact}
Ahmet~Onur Akman, Anastasia Psarou, Zolt{\'a}n~Gy{\"o}rgy Varga, Grzegorz Jamr{\'o}z, and Rafal Kucharski.
\newblock Impact of collective behaviors of autonomous vehicles on urban traffic dynamics: A multi-agent reinforcement learning approach.
\newblock In {\em Seventeenth European Workshop on Reinforcement Learning}, 2024.

\bibitem{RouteRL2024}
Ahmet~Onur Akman, Anastasia Psarou, Łukasz Gorczyca, Zoltán~György Varga, Grzegorz Jamróz, and Rafał Kucharski.
\newblock Routerl: Multi-agent reinforcement learning framework for urban route choice with autonomous vehicles, 2025.

\bibitem{arnott1990departure}
Richard Arnott, Andr{\'e} de~Palma, and Robin Lindsey.
\newblock Departure time and route choice for the morning commute.
\newblock {\em Transportation Research Part B: Methodological}, 24(3):209--228, 1990.

\bibitem{ault2021reinforcement}
James Ault and Guni Sharon.
\newblock Reinforcement learning benchmarks for traffic signal control.
\newblock In {\em Thirty-fifth Conference on Neural Information Processing Systems Datasets and Benchmarks Track}, 2021.

\bibitem{BAMDADMEHRABANI2024}
Behzad {Bamdad Mehrabani}, Jakob Erdmann, Luca Sgambi, Seyedehsan Seyedabrishami, and Maaike Snelder.
\newblock A multiclass simulation-based dynamic traffic assignment model for mixed traffic flow of connected and autonomous vehicles and human-driven vehicles.
\newblock {\em Transportmetrica A Transport Science}, 2024.

\bibitem{ben1999discrete}
Moshe Ben-Akiva and Michel Bierlaire.
\newblock Discrete choice methods and their applications to short term travel decisions.
\newblock In {\em Handbook of transportation science}, pages 5--33. Springer, 1999.

\bibitem{Bhat1999}
Chandra~R. Bhat and Frank~S. Koppelman.
\newblock {\em Activity-Based Modeling of Travel Demand}, pages 35--61.
\newblock Springer US, Boston, MA, 1999.

\bibitem{bie2010stability}
Jing Bie and Hong~K Lo.
\newblock Stability and attraction domains of traffic equilibria in a day-to-day dynamical system formulation.
\newblock {\em Transportation Research Part B: Methodological}, 44(1):90--107, 2010.

\bibitem{bou2023torchrl}
Albert Bou, Matteo Bettini, Sebastian Dittert, Vikash Kumar, Shagun Sodhani, Xiaomeng Yang, Gianni~De Fabritiis, and Vincent Moens.
\newblock Torchrl: A data-driven decision-making library for pytorch, 2023.

\bibitem{bovy2005modelling}
Piet~HL Bovy and Sascha Hoogendoorn-Lanser.
\newblock Modelling route choice behaviour in multi-modal transport networks.
\newblock {\em Transportation}, 32:341--368, 2005.

\bibitem{Cascetta}
Ennio Cascetta.
\newblock {\em Transportation System Analysis: Models and Applications}.
\newblock 01 2009.

\bibitem{chen2018exploring}
Peng Chen, Rui Tong, Guangquan Lu, and Yunpeng Wang.
\newblock Exploring travel time distribution and variability patterns using probe vehicle data: case study in beijing.
\newblock {\em Journal of Advanced Transportation}, 2018(1):3747632, 2018.

\bibitem{centralization}
Caroline Claus and Craig Boutilier.
\newblock The dynamics of reinforcement learning in cooperative multiagent systems.
\newblock In {\em Proceedings of the Fifteenth National/Tenth Conference on Artificial Intelligence/Innovative Applications of Artificial Intelligence}, AAAI '98/IAAI '98, page 746–752, USA, 1998. American Association for Artificial Intelligence.

\bibitem{correa2004selfish}
Jos{\'e}~R Correa, Andreas~S Schulz, and Nicol{\'a}s~E Stier-Moses.
\newblock Selfish routing in capacitated networks.
\newblock {\em Mathematics of Operations Research}, 29(4):961--976, 2004.

\bibitem{improved_learning_automata_route_choice}
Gabriel de~O.~Ramos and Ricardo Grunitzki.
\newblock An improved learning automata approach for the route choice problem.
\newblock In Fernando Koch, Felipe Meneguzzi, and Kiran Lakkaraju, editors, {\em Agent Technology for Intelligent Mobile Services and Smart Societies}, pages 56--67, Berlin, Heidelberg, 2015. Springer Berlin Heidelberg.

\bibitem{ippo}
Christian~Schr{\"{o}}der de~Witt, Tarun Gupta, Denys Makoviichuk, Viktor Makoviychuk, Philip H.~S. Torr, Mingfei Sun, and Shimon Whiteson.
\newblock Is independent learning all you need in the starcraft multi-agent challenge?
\newblock {\em CoRR}, abs/2011.09533, 2020.

\bibitem{di2017indifference}
Xuan Di, Henry~X Liu, Shanjiang Zhu, and David~M Levinson.
\newblock Indifference bands for boundedly rational route switching.
\newblock {\em Transportation}, 44:1169--1194, 2017.

\bibitem{stabilizing_experience_replay}
Jakob~N. Foerster, Nantas Nardelli, Gregory Farquhar, Philip H.~S. Torr, Pushmeet Kohli, and Shimon Whiteson.
\newblock Stabilising experience replay for deep multi-agent reinforcement learning.
\newblock {\em CoRR}, abs/1702.08887, 2017.

\bibitem{frejinger2009sampling}
Emma Frejinger, Michel Bierlaire, and Moshe Ben-Akiva.
\newblock Sampling of alternatives for route choice modeling.
\newblock {\em Transportation Research Part B: Methodological}, 43(10):984--994, 2009.

\bibitem{fukushima1984modified}
Masao Fukushima.
\newblock A modified frank-wolfe algorithm for solving the traffic assignment problem.
\newblock {\em Transportation Research Part B: Methodological}, 18(2):169--177, 1984.

\bibitem{Gonz_lez_2008}
Marta~C. González, César~A. Hidalgo, and Albert-László Barabási.
\newblock Understanding individual human mobility patterns.
\newblock {\em Nature}, 453(7196):779–782, June 2008.

\bibitem{indi_vs_diff_rewards_route_choice}
Ricardo Grunitzki, Gabriel de~Oliveira Ramos, and Ana Lucia~Cetertich Bazzan.
\newblock Individual versus difference rewards on reinforcement learning for route choice.
\newblock In {\em 2014 Brazilian Conference on Intelligent Systems}, pages 253--258, 2014.

\bibitem{sac}
Tuomas Haarnoja, Aurick Zhou, Pieter Abbeel, and Sergey Levine.
\newblock Soft actor-critic: Off-policy maximum entropy deep reinforcement learning with a stochastic actor, 2018.

\bibitem{DBLP:journals/corr/abs-1812-05905}
Tuomas Haarnoja, Aurick Zhou, Kristian Hartikainen, George Tucker, Sehoon Ha, Jie Tan, Vikash Kumar, Henry Zhu, Abhishek Gupta, Pieter Abbeel, and Sergey Levine.
\newblock Soft actor-critic algorithms and applications.
\newblock {\em CoRR}, abs/1812.05905, 2018.

\bibitem{he2012modeling}
Xiaozheng He and Henry~X Liu.
\newblock Modeling the day-to-day traffic evolution process after an unexpected network disruption.
\newblock {\em Transportation Research Part B: Methodological}, 46(1):50--71, 2012.

\bibitem{horni2016matsim}
Andreas Horni, Kai Nagel, and Kay~W. Axhausen, editors.
\newblock {\em The Multi-Agent Transport Simulation MATSim}.
\newblock Ubiquity Press, London, 2016.
\newblock License: CC-BY 4.0.

\bibitem{ejal}
Luis~A. Hurtado, Elena Mocanu, Phuong~H. Nguyen, Madeleine Gibescu, and René I.~G. Kamphuis.
\newblock Enabling cooperative behavior for building demand response based on extended joint action learning.
\newblock {\em IEEE Transactions on Industrial Informatics}, 14(1):127--136, 2018.

\bibitem{jamróz2024socialimpactcavs}
Grzegorz Jamróz, Ahmet~Onur Akman, Anastasia Psarou, Zoltán~Györgi Varga, and Rafał Kucharski.
\newblock Social impact of cavs -- coexistence of machines and humans in the context of route choice, 2024.

\bibitem{jiang2024blackbox}
Haozhe Jiang, Qiwen Cui, Zhihan Xiong, Maryam Fazel, and Simon~S. Du.
\newblock A black-box approach for non-stationary multi-agent reinforcement learning.
\newblock In {\em International Conference on Learning Representations (ICLR)}, 2024.

\bibitem{kopelias2020connected}
Pantelis Kopelias, Elissavet Demiridi, Konstantinos Vogiatzis, Alexandros Skabardonis, and Vassiliki Zafiropoulou.
\newblock Connected \& autonomous vehicles--environmental impacts--a review.
\newblock {\em Science of the total environment}, 712:135237, 2020.

\bibitem{kucharski2017estimating}
Rafał Kucharski and Arkadiusz Drabicki.
\newblock Estimating macroscopic volume delay functions with the traffic density derived from measured speeds and flows.
\newblock {\em Journal of Advanced Transportation}, 2017(1):4629792, 2017.

\bibitem{lazar2021learningdynamicallyrouteautonomous}
Daniel~A. Lazar, Erdem Bıyık, Dorsa Sadigh, and Ramtin Pedarsani.
\newblock Learning how to dynamically route autonomous vehicles on shared roads, 2021.

\bibitem{lobo_intas_2020}
Silas~C. Lobo, Stefan Neumeier, Evelio M.~G. Fernandez, and Christian Facchi.
\newblock Intas -- the ingolstadt traffic scenario for sumo, 2020.

\bibitem{SUMO2018}
Pablo~Alvarez Lopez, Michael Behrisch, Laura Bieker-Walz, Jakob Erdmann, Yun-Pang Fl{\"o}tter{\"o}d, Robert Hilbrich, Leonhard L{\"u}cken, Johannes Rummel, Peter Wagner, and Evamarie Wie{\ss}ner.
\newblock Microscopic traffic simulation using sumo.
\newblock In {\em The 21st IEEE International Conference on Intelligent Transportation Systems}. IEEE, 2018.

\bibitem{marl_actor_critic_mixed_cooperative_competitive_envs}
Ryan Lowe, Yi~Wu, Aviv Tamar, Jean Harb, Pieter Abbeel, and Igor Mordatch.
\newblock Multi-agent actor-critic for mixed cooperative-competitive environments.
\newblock {\em CoRR}, abs/1706.02275, 2017.

\bibitem{lowe2020multiagentactorcriticmixedcooperativecompetitive}
Ryan Lowe, Yi~Wu, Aviv Tamar, Jean Harb, Pieter Abbeel, and Igor Mordatch.
\newblock Multi-agent actor-critic for mixed cooperative-competitive environments, 2020.

\bibitem{lu2024centralized}
C.~Lu, Q.~Bao, S.~Xia, et~al.
\newblock Centralized reinforcement learning for multi-agent cooperative environments.
\newblock {\em Evol. Intel.}, 17(2):267--273, 2024.

\bibitem{Ma2020FeudalMD}
Jinming Ma and Feng Wu.
\newblock Feudal multi-agent deep reinforcement learning for traffic signal control.
\newblock In {\em Adaptive Agents and Multi-Agent Systems}, 2020.

\bibitem{NIPS2009_3a15c7d0}
Hamid Maei, Csaba Szepesv\'{a}ri, Shalabh Bhatnagar, Doina Precup, David Silver, and Richard~S Sutton.
\newblock Convergent temporal-difference learning with arbitrary smooth function approximation.
\newblock In Y.~Bengio, D.~Schuurmans, J.~Lafferty, C.~Williams, and A.~Culotta, editors, {\em Advances in Neural Information Processing Systems}, volume~22. Curran Associates, Inc., 2009.

\bibitem{maven}
Anuj Mahajan, Tabish Rashid, Mikayel Samvelyan, and Shimon Whiteson.
\newblock {MAVEN:} multi-agent variational exploration.
\newblock {\em CoRR}, abs/1910.07483, 2019.

\bibitem{mei2023macpomultiagentexperiencereplay}
Yongsheng Mei, Hanhan Zhou, Tian Lan, Guru Venkataramani, and Peng Wei.
\newblock Mac-po: Multi-agent experience replay via collective priority optimization, 2023.

\bibitem{merchant1978optimality}
Deepak~K Merchant and George~L Nemhauser.
\newblock Optimality conditions for a dynamic traffic assignment model.
\newblock {\em Transportation Science}, 12(3):200--207, 1978.

\bibitem{nair2021awacacceleratingonlinereinforcement}
Ashvin Nair, Abhishek Gupta, Murtaza Dalal, and Sergey Levine.
\newblock Awac: Accelerating online reinforcement learning with offline datasets, 2021.

\bibitem{nayak2021autonomous}
Biraja~Prasad Nayak, Lopamudra Hota, Arun Kumar, Ashok~Kumar Turuk, and Peter~HJ Chong.
\newblock Autonomous vehicles: Resource allocation, security, and data privacy.
\newblock {\em IEEE Transactions on Green Communications and Networking}, 6(1):117--131, 2021.

\bibitem{neun2023traffic4cast}
Moritz Neun, Christian Eichenberger, Henry Martin, et~al.
\newblock Traffic4cast at neurips 2022 -- predict dynamics along graph edges from sparse node data: Whole city traffic and eta from stationary vehicle detectors.
\newblock {\em arXiv preprint arXiv:2303.07758}, 2023.

\bibitem{papoudakis2021benchmarkingmultiagentdeepreinforcement}
Georgios Papoudakis, Filippos Christianos, Lukas Schäfer, and Stefano~V. Albrecht.
\newblock Benchmarking multi-agent deep reinforcement learning algorithms in cooperative tasks, 2021.

\bibitem{influence_of_AVs_on_car_following_behavior}
Yalda Rahmati, Mohammadreza Khajeh~Hosseini, Alireza Talebpour, Benjamin Swain, and Christopher Nelson.
\newblock Influence of autonomous vehicles on car-following behavior of human drivers.
\newblock {\em Transportation Research Record: Journal of the Transportation Research Board}, 2673:036119811986262, 07 2019.

\bibitem{regret_route_choice}
Gabriel Ramos, Ana Bazzan, and Bruno da~Silva.
\newblock Analysing the impact of travel information for minimising the regret of route choice.
\newblock {\em Transportation Research Part C: Emerging Technologies}, 88:257--271, 03 2018.

\bibitem{qmix}
Tabish Rashid, Mikayel Samvelyan, Christian~Schr{\"{o}}der de~Witt, Gregory Farquhar, Jakob~N. Foerster, and Shimon Whiteson.
\newblock {QMIX:} monotonic value function factorisation for deep multi-agent reinforcement learning.
\newblock {\em CoRR}, abs/1803.11485, 2018.

\bibitem{goldman2024partiallyautonomous}
Goldman Sachs.
\newblock Partially autonomous cars forecast to comprise 10 percent of new vehicle sales by 2030, 2024.
\newblock Accessed: 2025-01-27.

\bibitem{ppo}
John Schulman, Filip Wolski, Prafulla Dhariwal, Alec Radford, and Oleg Klimov.
\newblock Proximal policy optimization algorithms, 2017.

\bibitem{schwarting2019social}
Wilko Schwarting, Alyssa Pierson, Javier Alonso-Mora, Sertac Karaman, and Daniela Rus.
\newblock Social behavior for autonomous vehicles.
\newblock {\em Proceedings of the National Academy of Sciences}, 116(50):24972--24978, 2019.

\bibitem{vdn}
Peter Sunehag, Guy Lever, Audrunas Gruslys, Wojciech~Marian Czarnecki, Vin{\'{\i}}cius~Flores Zambaldi, Max Jaderberg, Marc Lanctot, Nicolas Sonnerat, Joel~Z. Leibo, Karl Tuyls, and Thore Graepel.
\newblock Value-decomposition networks for cooperative multi-agent learning.
\newblock {\em CoRR}, abs/1706.05296, 2017.

\bibitem{taiebat2018review}
Morteza Taiebat, Austin~L Brown, Hannah~R Safford, Shen Qu, and Ming Xu.
\newblock A review on energy, environmental, and sustainability implications of connected and automated vehicles.
\newblock {\em Environmental science \& technology}, 52(20):11449--11465, 2018.

\bibitem{talebpour2016influence}
Alireza Talebpour and Hani~S Mahmassani.
\newblock Influence of connected and autonomous vehicles on traffic flow stability and throughput.
\newblock {\em Transportation research part C: emerging technologies}, 71:143--163, 2016.

\bibitem{iql}
Ming Tan.
\newblock {\em Multi-agent reinforcement learning: independent vs. cooperative agents}, page 487–494.
\newblock Morgan Kaufmann Publishers Inc., San Francisco, CA, USA, 1997.

\bibitem{terry2021pettingzoo}
J~Terry, Benjamin Black, Nathaniel Grammel, Mario Jayakumar, Ananth Hari, Ryan Sullivan, Luis~S Santos, Clemens Dieffendahl, Caroline Horsch, Rodrigo Perez-Vicente, et~al.
\newblock Pettingzoo: Gym for multi-agent reinforcement learning.
\newblock {\em Advances in Neural Information Processing Systems}, 34:15032--15043, 2021.

\bibitem{Thomasini+2023}
Luiz~A. Thomasini, Lucas~N. Alegre, Gabriel~O. Ramos, and Ana L.~C. Bazzan.
\newblock Routechoiceenv: a route choice library for multiagent reinforcement learning.
\newblock In {\em Adaptive and Learning Agents Workshop at AAMAS}, 2023.

\bibitem{Treiber_2000}
Martin Treiber, Ansgar Hennecke, and Dirk Helbing.
\newblock Congested traffic states in empirical observations and microscopic simulations.
\newblock {\em Physical Review E}, 62(2):1805–1824, August 2000.

\bibitem{AgentRewardShaping}
Kagan Tumer and Adrian Agogino.
\newblock Agent reward shaping for alleviating traffic congestion.
\newblock 01 2006.

\bibitem{turocy2001gambit}
Theodore~L. Turocy.
\newblock Gambit: Software tools for game theory, version 0.2007.01.30.
\newblock Technical Report 01-01, Texas A\&M University Department of Economics, 2001.

\bibitem{Wardrop1952ROADPS}
J.~G. Wardrop.
\newblock Road paper. some theoretical aspects of road traffic research.
\newblock 1952.

\bibitem{watling2013modelling}
David~P Watling and Giulio~E Cantarella.
\newblock Modelling sources of variation in transportation systems: theoretical foundations of day-to-day dynamic models.
\newblock {\em Transportmetrica B: Transport Dynamics}, 1(1):3--32, 2013.

\bibitem{yu2022surprisingeffectivenessppocooperative}
Chao Yu, Akash Velu, Eugene Vinitsky, Jiaxuan Gao, Yu~Wang, Alexandre Bayen, and Yi~Wu.
\newblock The surprising effectiveness of ppo in cooperative, multi-agent games, 2022.

\bibitem{mappo}
Chao Yu, Akash Velu, Eugene Vinitsky, Yu~Wang, Alexandre~M. Bayen, and Yi~Wu.
\newblock The surprising effectiveness of {MAPPO} in cooperative, multi-agent games.
\newblock {\em CoRR}, abs/2103.01955, 2021.

\bibitem{drones7030150}
Longfei Yue, Rennong Yang, Jialiang Zuo, Mengda Yan, Xiaoru Zhao, and Maolong Lv.
\newblock Factored multi-agent soft actor-critic for cooperative multi-target tracking of uav swarms.
\newblock {\em Drones}, 7(3), 2023.

\bibitem{zhao2020sim}
Wenshuai Zhao, Jorge~Pe{\~n}a Queralta, and Tomi Westerlund.
\newblock Sim-to-real transfer in deep reinforcement learning for robotics: a survey.
\newblock In {\em 2020 IEEE symposium series on computational intelligence (SSCI)}, pages 737--744. IEEE, 2020.

\bibitem{ZHOU2020124895}
Bo~Zhou, Qiankun Song, Zhenjiang Zhao, and Tangzhi Liu.
\newblock A reinforcement learning scheme for the equilibrium of the in-vehicle route choice problem based on congestion game.
\newblock {\em Applied Mathematics and Computation}, 371:124895, 2020.

\bibitem{zhou2024modeling}
Wenhan Zhou, Jiancheng Weng, Tongfei Li, Bo~Fan, and Yang Bian.
\newblock Modeling the road network capacity in a mixed hv and cav environment.
\newblock {\em Physica A: Statistical Mechanics and its Applications}, 636:129526, 2024.

\end{thebibliography}

\newpage
\appendix

\section{Experiments}
\label{sec:experiments_details}
The experiments were conducted using version 0.0.1 of the RouteRL framework \cite{RouteRL2024} for the \texttt{TRY} network and version 1.0.0 for the Ingolstadt network. RouteRL is released under the MIT License. 
All experiments on the \texttt{TRY} network 
involve 200 policy updates, with each update consisting of 40 frames collected per agent (interpreted as episodes or days). Training is performed for 10 epochs, using a minibatch size of 2, and the Tanh activation function. The algorithm scripts are derived from the state-of-the-art (SOTA) implementations provided by the TorchRL library \cite{bou2023torchrl}. All experiments were conducted using version 0.3.0 of the open-source TorchRL library to ensure reproducibility.

In the experiment conducted on the Ingolstadt network, 4,000 frames were used with 800 policy updates. The AVs were introduced into the traffic network from episode 500. The training was performed using 1 epoch and a minibatch size of 32. All hyperparameters were consistent with those listed in Table \ref{tab:hyperparameters}.

\begin{figure}[H]
\vskip 0.1in
\begin{center}
\centerline{\includegraphics[width=1\linewidth]{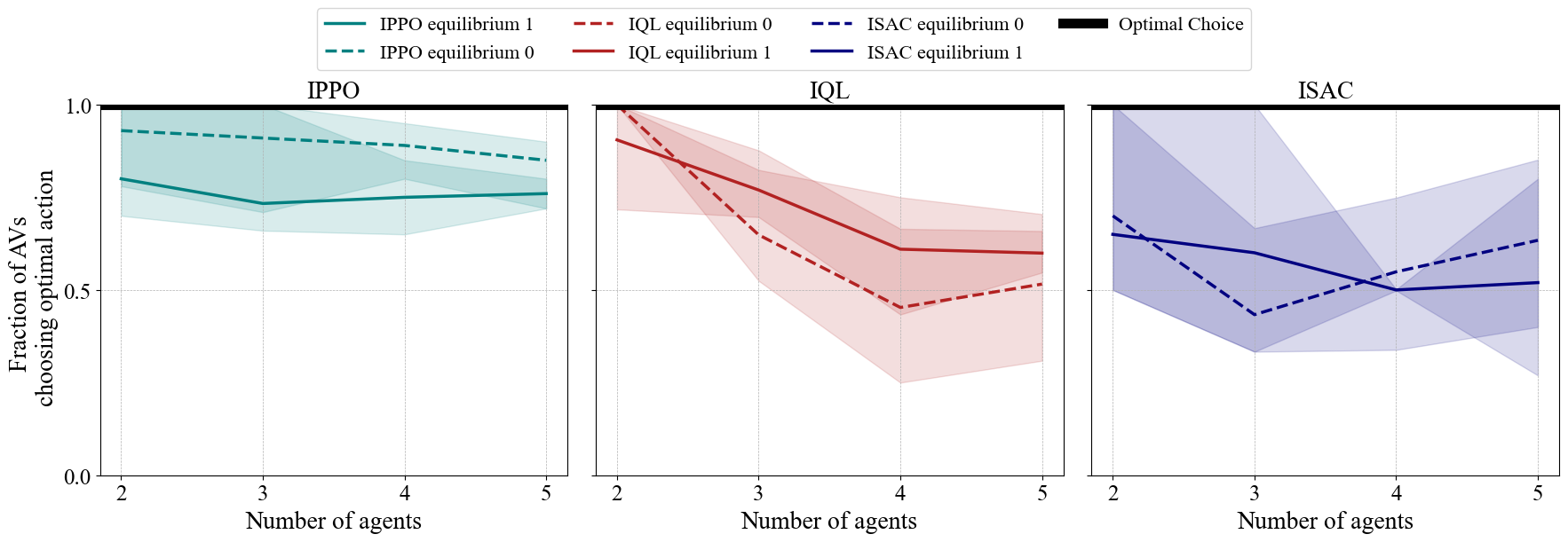}}
\caption{With an increasing number of agents, a fraction of those, who learn the optimal policy decreases, achieving similar efficiency to random choice (equilibrium 0: system optimal - user equilibrium - Figure \ref{fig:overview}a, equilibrium 1: suboptimal user equilibrium - Figure \ref{fig:overview}b). This appears to happen with as few as four agents for some of the tested algorithms (ISAC, IQL) that failed to converge in Figure \ref{fig:marl-proportions}. See Appendix \ref{sec:experiments_details} for the experiments' setup.}
\label{fig:different-fleet-number}
\end{center}
\vskip -0.1in
\end{figure}

\begin{figure}[H]
\vskip 0.1in
\begin{center}
\centerline{\includegraphics[width=0.5\linewidth]{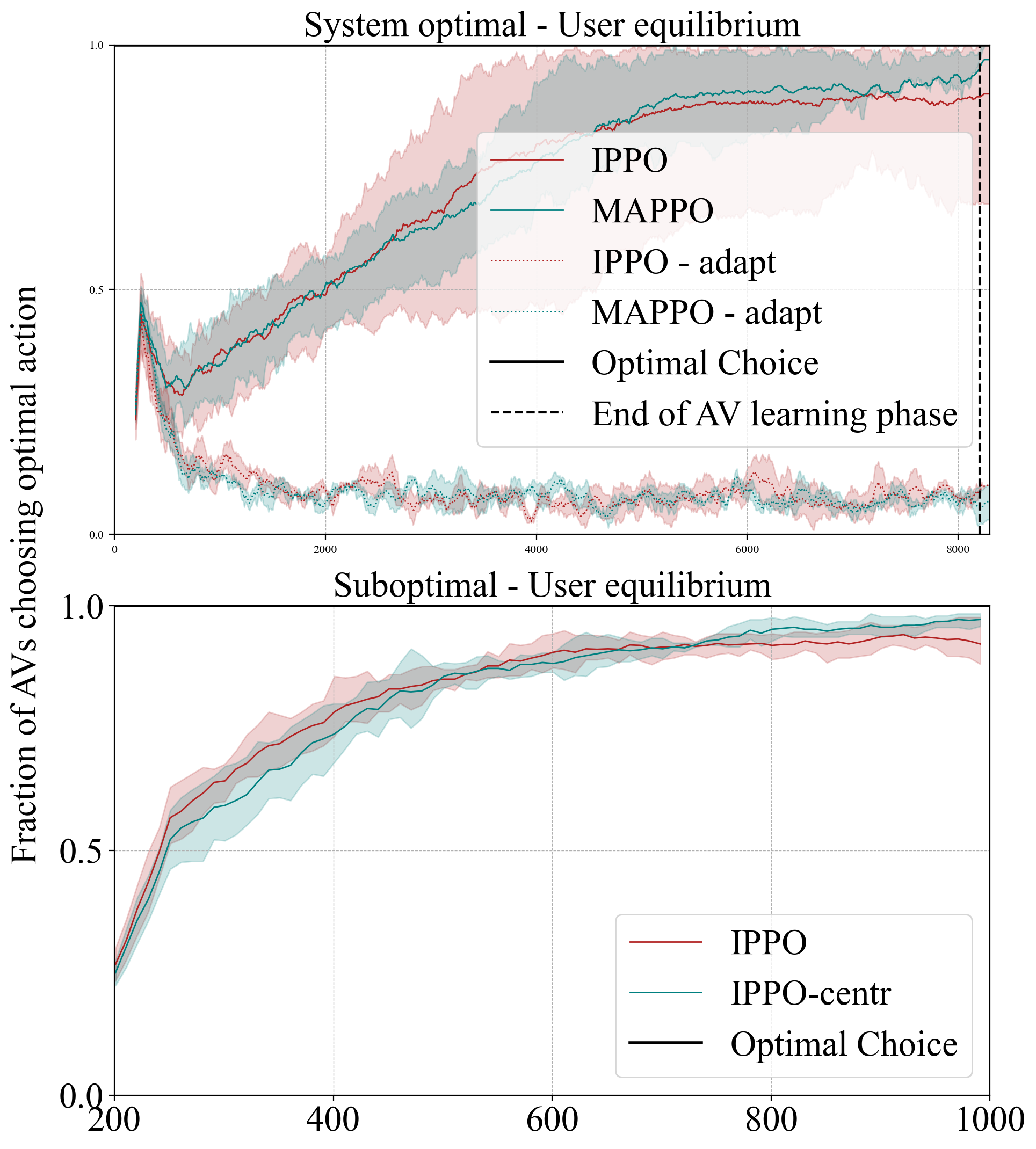}}
\caption{\textbf{Top}. IPPO and MAPPO still converge if the adaptation term is added to the model (dotted lines). However, the system diverts from the optimal state to the suboptimal, where the total costs are higher. {\textbf{Bottom}.} The centralized version of IPPO can, in some cases, accelerate convergence reducing the time required to reach an optimal solution, see Appendix \ref{Appendix:exceptions-centralized} for exceptions. However, this reduction is limited to a few days. See \ref{sec:experiments_details} for the experiments' setup.}
\label{fig:adapt-centr}
\end{center}
\vskip -0.1in
\end{figure}

Each experiment was replicated a different number of times, depending on the scenario. In the case of multiple simultaneously routing AVs (described in Section \ref{sec:multi-avs} and shown in Figure \ref{fig:marl-proportions}(a, b)), each algorithm experiment had 10 replications for each equilibrium. For the experiments investigating the minimum number of simultaneously routing AVs that can destabilize the traffic network (described in Section \ref{para:fleet_size} and shown in Figure \ref{fig:different-fleet-number}), each algorithm was replicated for five independent runs for each equilibrium. The experiments exploring the effects of non-determinism and centralization (described in Sections \ref{sec:non-determinism}, \ref{sec:centr} and shown in Figure \ref{fig:marl-proportions}c and \ref{fig:adapt-centr}) were conducted for 3 independent replications per setting.
Hyperparameter optimization was performed for each algorithm and setting (e.g., centralization) 
separately.

\begin{table}[H]
\caption{Hyperparameters used in the experiments.}
\label{tab:hyperparameters}
\vskip 0.15in
\begin{center}
\begin{small}
\begin{sc}
\resizebox{\textwidth}{!}{

    \begin{tabular}{lccccccccc}
    \toprule
    Hyperparameter & IQL & VDN & QMIX & MASAC & ISAC & MAPPO & IPPO & IPPO/MAPPO - Centr\\
    \midrule
    Learning rate       & 1e-4 & 1e-2 & 1e-3 & 1e-5 & 1e-5 & 1e-5 & 1e-5 & 1e-4\\
    Memory size         & 1600 & 1600 & 1600 & 1600 & 1600 & -    & -   & - \\
    Max gradient norm   & 1e-3 & 2    & 1.5  & 0.5  & 0.5  & 0.5  & 0.5 & 1.0 \\
    Tau                 & 1    & 1e-2 & 1e-2 & 5e-3 & 5e-3 & -    & -   & - \\
    Gamma               & 0.9  & 0.85 & 0.95 & 0.98 & 0.99 & 0.99 & 0.99 & 0.85 \\
    Lambda              & -    & -    & -    & -    & -    & 1    & 1 & 0.9\\
    Clip epsilon        & -    & -    & -    & -    & -    & 0.2  & 0.2 & 0.2  \\
    Entropy coefficient & -    & -    & -    & -    & -    & 1e-4 & 1e-4 & 1e-3\\
    \bottomrule
    \end{tabular}
}
\end{sc}
\end{small}
\end{center}
\vskip -0.1in
\end{table}

\paragraph{Hardware.} Our experiments were carried out on our institution's computing nodes with resources allocated as listed in Table \ref{tab:hardware}.

\begin{table}[H]
\caption{Summary of the hardware used for the experiments.}
\label{tab:hardware}
\vskip 0.15in
\begin{center}
\begin{footnotesize}
\begin{sc}
\begin{tabular}{@{}ll@{}}
    \toprule
    Component & Specification \\
    \midrule
    CPU                    & Intel(R) Xeon(R) Gold 5122 CPU, 3.60GHz \\
    GPU                    & NVIDIA GeForce RTX 2080 \\
    RAM                    & 40 GB allocated per job \\
    Operating system       & Ubuntu 24.04.1 LTS \\
    Job scheduler          & SLURM \\
    SUMO version           & 1.18.0 \\
    \bottomrule
\end{tabular}
\end{sc}
\end{footnotesize}
\end{center}
\vskip -0.1in
\end{table}

\textbf{Execution time.} Experiment compute time depends on the MARL algorithm. We share the computation time of some representative cases in Table \ref{tab:comp_time}.

\begin{table}[H]
\caption{Computation time of representative experiments.}
\label{tab:comp_time}
\vskip 0.15in
\begin{center}
\begin{footnotesize}
\begin{sc}
\begin{tabular}{@{}llc@{}}
\toprule
Traffic network       & Algorithm & Runtime (min) \\
\midrule
TRY     & IQL                    & $\sim$64\\
TRY   & QMIX                    & $\sim$80 \\
TRY   & VDN                    & $\sim$80 \\
TRY        & IPPO                    & $\sim$200   \\
TRY        & MAPPO                    & $\sim$200   \\
TRY        & ISAC                    & $\sim$220   \\
TRY        & MASAC                  & $\sim$220   \\
Ingolstadt     & IPPO                    & $\sim$7200 (120 hrs)\\
\bottomrule
\end{tabular}
\end{sc}
\end{footnotesize}
\end{center}
\vskip -0.1in

\end{table}

Overall, the experiments described in the paper required approximately 2,000 hours of computation. Full research required up to twice as much computation in total, due to preliminary testing, errors, and our curiosity-driven exploration.

\section{Plots}
\label{sec:plot_description}

In all plots, except Figure \ref{fig:ingolstadt-ippo}, the lines represent the mean across multiple replications for each episode, and the error bars indicate the 15th and 85th percentile values. In Figure \ref{fig:single-agent}, the line depicts the mean across replications, while the error band accounts for variations due to different agent start times and the use of various algorithms, as described in Section \ref{sec:single_agents}. All plotted data smoothed using a moving average method with a window size of value 10.

\section{SUMO}
\label{sec:SUMO_appendix}

SUMO is the traffic simulator used in this study and is licensed under the Eclipse Public License Version 2.0 (EPL V2). Throughout our experiments, SUMO operates under deterministic conditions, except Figure \ref{fig:marl-proportions}c, where we explicitly introduce non-determinism. Under deterministic settings, if all vehicles select the same routes across consecutive iterations, their travel times remain identical across runs. 

In the SUMO screenshots below, red vehicles represent human drivers, and yellow vehicles depict AVs. The SUMO network file used to simulate the \texttt{TRY} network is provided in the supplementary material, along with simulation videos.

\begin{figure}[ht]
    \centering
    \begin{subfigure}[b]{\textwidth}
        \centering
        \includegraphics[width=\linewidth]{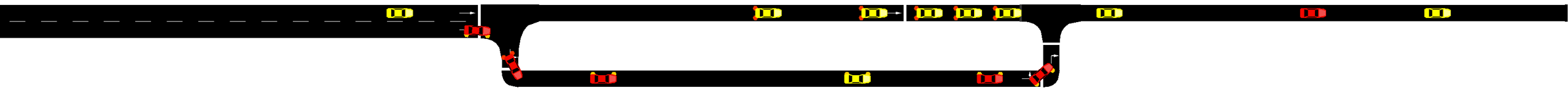}
        \caption{System optimal user equilibrium (see Figure \ref{fig:overview})a.}
        \label{fig:figure2}
    \end{subfigure}
    \hfill
    \hfill
    \begin{subfigure}[b]{\textwidth}
        \centering
        \includegraphics[width=\linewidth]{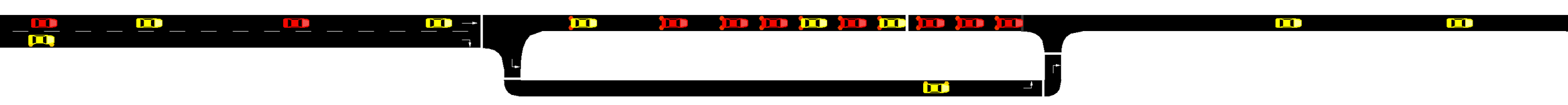}
        \caption{Suboptimal user equilibrium (see Figure \ref{fig:overview})b.}
        \label{fig:figure1}
    \end{subfigure}
    \caption{SUMO screenshots illustrate the network under the two distinct equilibria. }
    \label{fig:sumo_screenshots}
\end{figure}

\begin{figure}[ht]
    \centering
    \includegraphics[width=0.7\linewidth]{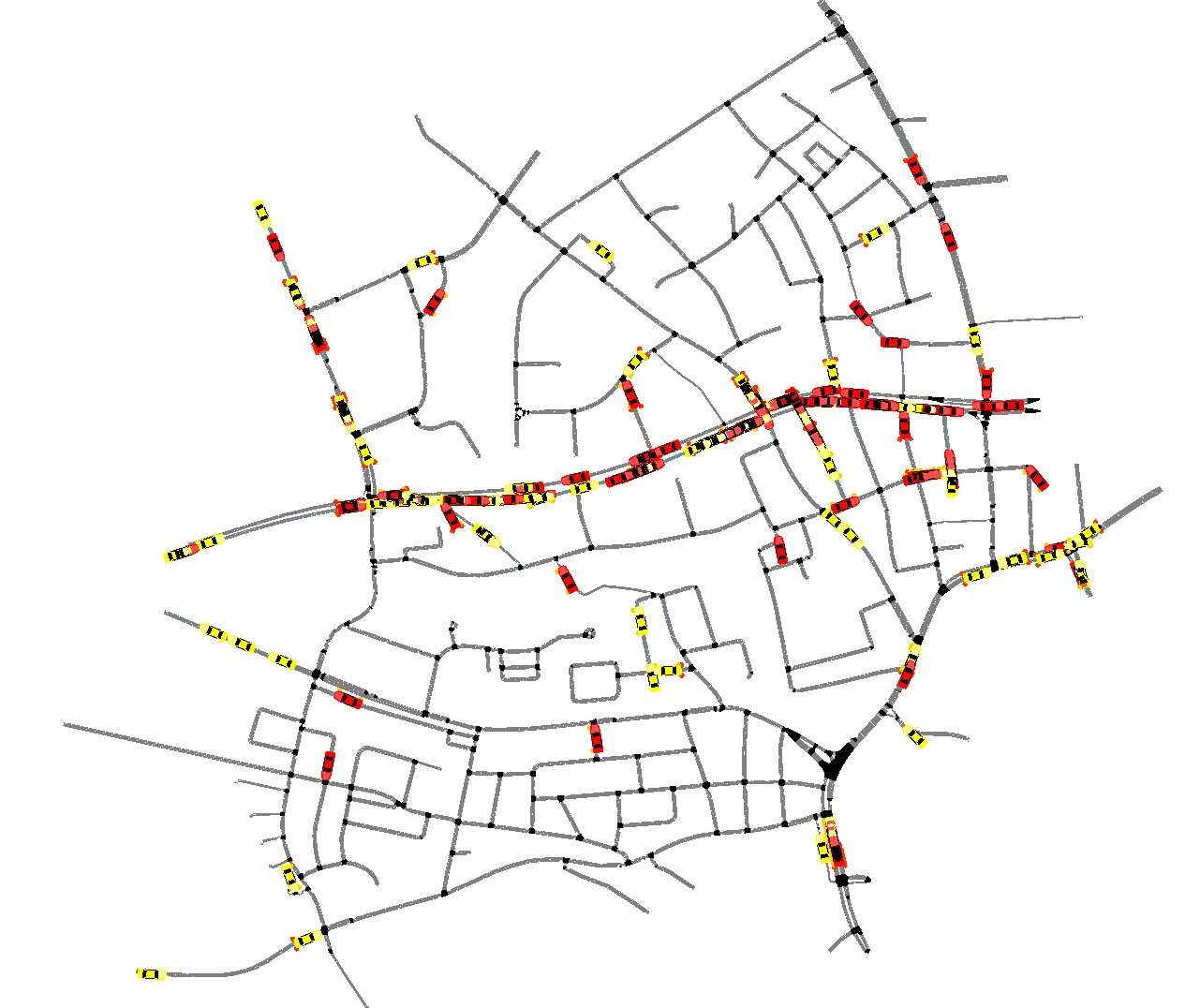}
    \caption{Sumo screenshot from the Ingolstadt network.}
    \label{fig:ingolstadt_network}
\end{figure}

\section{Ingolstadt network experiment}
\label{sec:ingolstadt_network}

The Ingolstadt network consists of 850 links, with demand data sourced from InTAS \cite{lobo_intas_2020} and used in the RESCO benchmark \cite{Ma2020FeudalMD}. This demand includes 1,000 agents. In our experiment, we consider that 400 human agents transitioned to using AVs in episode 500. Each agent learns to select the optimal path from four available alternatives. These alternative paths are generated using \texttt{Janux} software \cite{Janux}, which is integrated into the RouteRL framework. 

As shown in Figure \ref{fig:ingolstadt-ippo}, initially, when there are only humans, the system remains stable, exhibiting constant average travel time. However, starting at episode 500, the introduction of AVs destabilizes the system, leading to oscillatory and increased patterns in travel times for human agents.

\section{Brief theoretical analysis of MARL convergence}
\label{sec:theoretical_analysis}

This section outlines the theoretical barriers to the convergence of the MARL algorithms employed in this paper. Although empirical performance is often emphasized in MARL, for example in \cite{ippo, mei2023macpomultiagentexperiencereplay}, theoretical convergence guarantees are often lacking. Moreover, in recent benchmarks, agents are trained for millions of timesteps \cite{ippo, papoudakis2021benchmarkingmultiagentdeepreinforcement}.

In IL methods, such as IQL, each agent treats others as part of the environment. This breaks the Markov assumption and leads to non-stationarity, as the transition dynamics $P(s'|s, a_i)$ are no longer fixed, due to the evolving policies of other agents. As a result, even though IQL is a baseline used in MARL problems that often works well in practice, it lacks theoretical convergence guarantees \cite{papoudakis2021benchmarkingmultiagentdeepreinforcement, stabilizing_experience_replay, marl_actor_critic_mixed_cooperative_competitive_envs}.


Value factorization algorithms, such as VDN and QMIX assume that the joint action-value function $Q_{tot}(s,a)$ can be decomposed into individual functions $Q_{i}(s, a_i)$, either additively (VDN) or under monotonicity constraint (QMIX). This assumption is violated in many coordination tasks, especially those evolving non-monotonic utility landscapes \cite{maven}. 
However, this property does not guarantee non-convergence.


As far as the SAC algorithm is concerned prior work has mentioned that MASAC cannot scale easily to the large joint action spaces inherent in multi-agent settings \cite{drones7030150}. However, to the best of our knowledge, there is no proof to support this claim.


\section{Cumulative time difference}
\label{sec:cumulative_time_differences}

\begin{table}[H]
\caption{\textbf{Cumulative travel time differences} (in hours) with standard deviations between experiment replications for each user equilibrium and algorithm. The lowest cumulative travel time differences are highlighted in \textbf{bold}.}
\label{tab:cumtimedif}
\begin{center}
\begin{small}
\begin{sc}
\begin{tabular}{@{}llcc@{}}
\toprule
\multicolumn{1}{l|}{Algorithm} & \multicolumn{1}{c}{System Optimum \& User Equilibrium} & \multicolumn{1}{c}{Suboptimal \& User Equilibrium} \\ 
\midrule
\multicolumn{1}{l|}{IPPO} & \multicolumn{1}{c|}{12.87 (7.17)} & -0.21 (0.09) \\
\multicolumn{1}{l|}{MAPPO} & \multicolumn{1}{c|}{\textbf{11.80 (3.72)}}  & -0.19 (0.16) \\
\multicolumn{1}{l|}{ISAC} & \multicolumn{1}{c|}{33.33 (0.19)} & -4.77 (0.04) \\
\multicolumn{1}{l|}{MASAC} & \multicolumn{1}{c|}{33.27 (0.39)} & \textbf{-4.78 (0.04)} \\
\multicolumn{1}{l|}{QMIX} & \multicolumn{1}{c|}{36.20 (12.66)}  & -3.95 (1.47) \\
\multicolumn{1}{l|}{VDN} & \multicolumn{1}{c|}{32.88 (0.92)} & -4.72 (0.07) \\
\multicolumn{1}{l|}{IQL} & \multicolumn{1}{c|}{33.67 (1.53)} & -3.94 (0.30) \\ 
\bottomrule
\end{tabular}
\end{sc}
\end{small}
\end{center}
\vskip -0.1in
\end{table}
Positive cumulative travel time differences, i.e. \(c_t > 0\), mean that on average, human agents after mutation experience longer travel times, thus the impact of AVs introduction on them is negative. This occurred in the system's optimal user equilibrium scenario for each algorithm tested. In the suboptimal user equilibrium, \(c_t < 0\) for each algorithm. This means that human agents decrease their travel times compared to the ones they were experiencing before the introduction of AVs. This can be attributed to the fact that humans follow the route that has priority (route 1) and these values are negligible.

Let \(\mathcal{H}\) be the set of all agents that do not mutate to AVs. In addition, let \(Ex\) denote the number of experiment replications and \(Eps\) the number of episodes from the moment of mutation till the end of each experiment. For each \(i\in Ex\), \(e\in Eps\) and agent \(h\in\mathcal{H}\), we annotate travel time by \(t^i_{e,h}\). Let \(\bar{t}^i_e\) describe the average travel time of agents from \(\mathcal{H}\) for each episode \(e \in Eps\) and experiment \(i\in Ex\):
\[
    \bar{t}^i_e = \frac{1}{|\mathcal{H}||Ex|}\sum_{h\in\mathcal{H}}t^i_{e,h}.
\]
For each MARL algorithm and user equilibrium, the cumulative travel time difference \(c_t\) is given by the following formula:
\[
    c_t =\sum_{i\in Ex} \sum_{e\in Eps} \left( \bar{t}_e^i - \bar{t}_{\text{lh}}^i\right),
\]
where \(\bar{t}_{\text{lh}}^{i}\) stands for the average travel time of agents \(\mathcal{H}\) in the last episode before the mutation in the \(i\)-th experiment, i.e. 
\[
    \bar{t}^i_{\text{lh}} = \frac{1}{|\mathcal{H}||Ex|}\sum_{h\in\mathcal{H}}t^i_{\text{lh},h},
\]
where lh stands for the last episode before introducing AVs. For each algorithm and each equilibrium, we conducted 10 experiment replications. The set \(\mathcal{H}\) contains 12 human agents. Mutation began in the 201-st episode and experiments continued until episode 8300. The values in the \ref{tab:cumtimedif} are cumulative travel time differences and the standard deviations over experiments.

\section{Critical fleet analysis}
\label{sec:critical_fleet_analyxix_appendix}

\section{Centralization and adaptation}
\label{sec:centr-adapt-appendix}

\section{Limits of the centralized case}

\label{Appendix:exceptions-centralized}
\begin{figure}[H]
\vskip 0.2in
\begin{center}
\centerline{\includegraphics[width=0.6\linewidth]{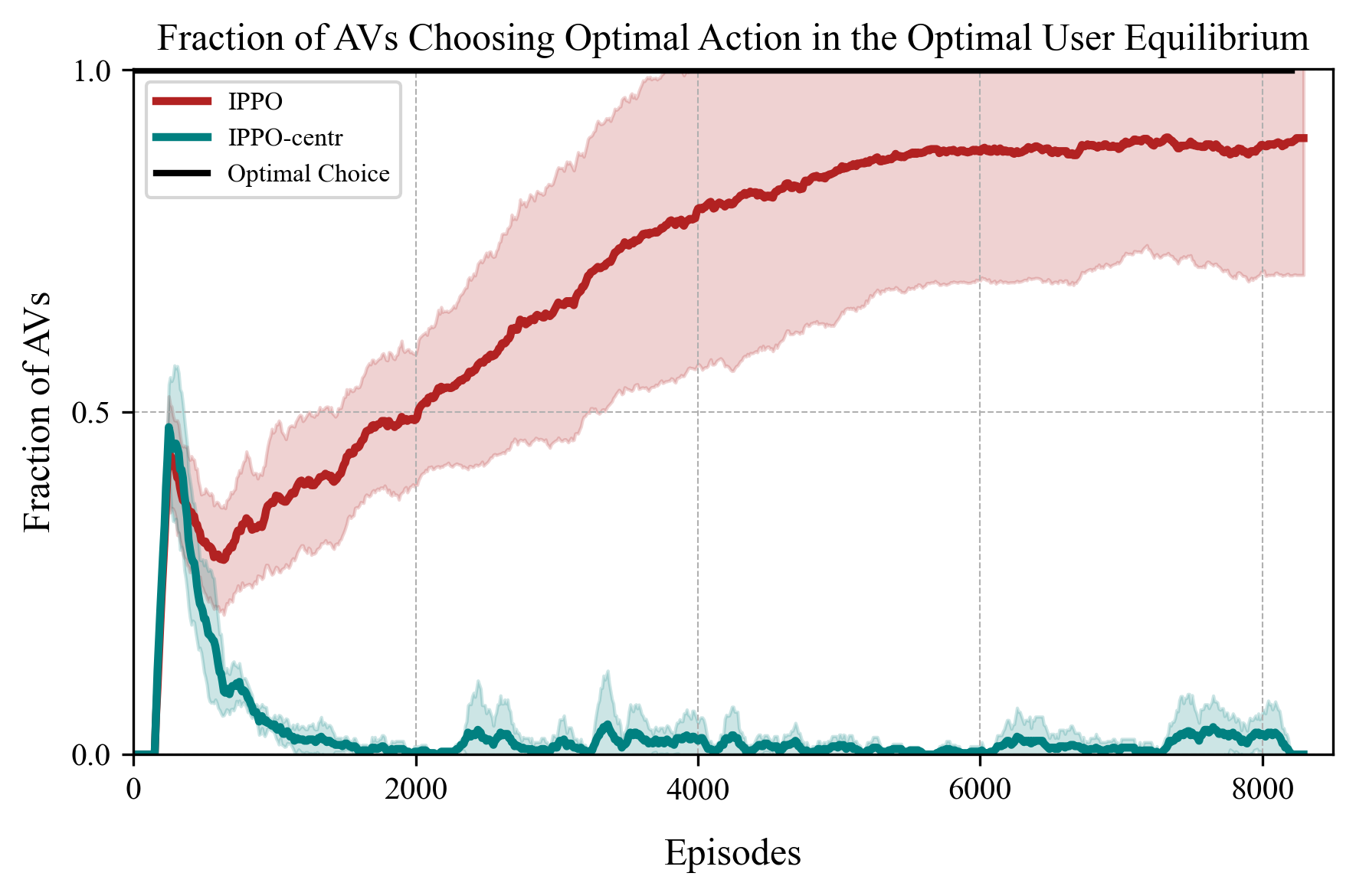}}
\caption{In the system optimal user equilibrium case (\ref{fig:overview}a), the centralized IPPO converges to the suboptimal solution, showing the limitations of the centralized case. This contrasts with the result shown in Figure \ref{fig:adapt-centr} bottom, where the centralized case converges to the optimal solution faster than the decentralized version of the algorithm.}
\label{fig:centralized-cases}
\end{center}
\vskip -0.2in
\end{figure}

\section{Travel times}
\begin{figure}[H]
\vskip 0.2in
\begin{center}
\centerline{\includegraphics[width=\textwidth]{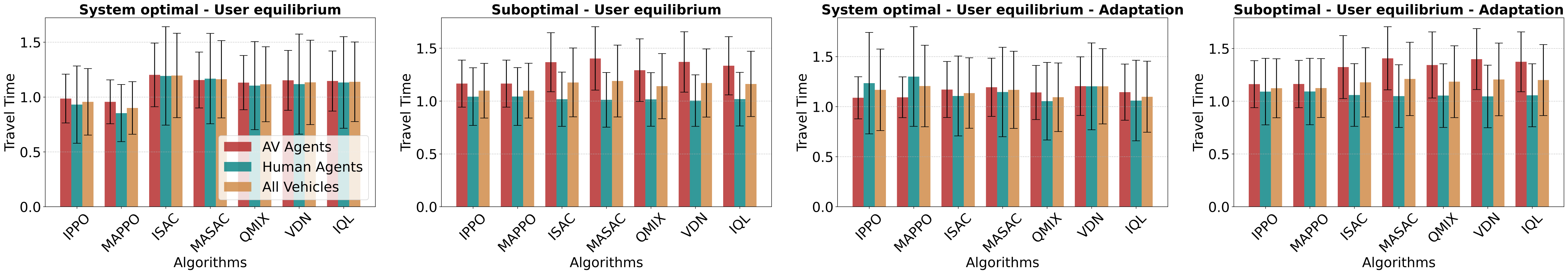}}
\caption{\textbf{Mean travel times} of AVs, human agents, and both in the simulation. The two plots on the left show that IPPO and MAPPO algorithms achieve shorter travel times for both AVs and human agents, as they converge to more optimal solutions, according to Figure \ref{fig:marl-proportions}(a, b). In the two plots on the right, that represent the scenario where humans adapt their options (\ref{par:human_agents}), AVs achieve smaller travel times using the same algorithms. Human drivers gain no benefit. The error bars represent the standard deviation between three distinct experiments conducted for each algorithm in each equilibrium.}
\label{fig:travel_times}
\end{center}
\vskip -0.2in
\end{figure}

\section{Adaptation}
\begin{figure}[H]
\vskip 0.1in
\begin{center}
\centerline{\includegraphics[width=\columnwidth]{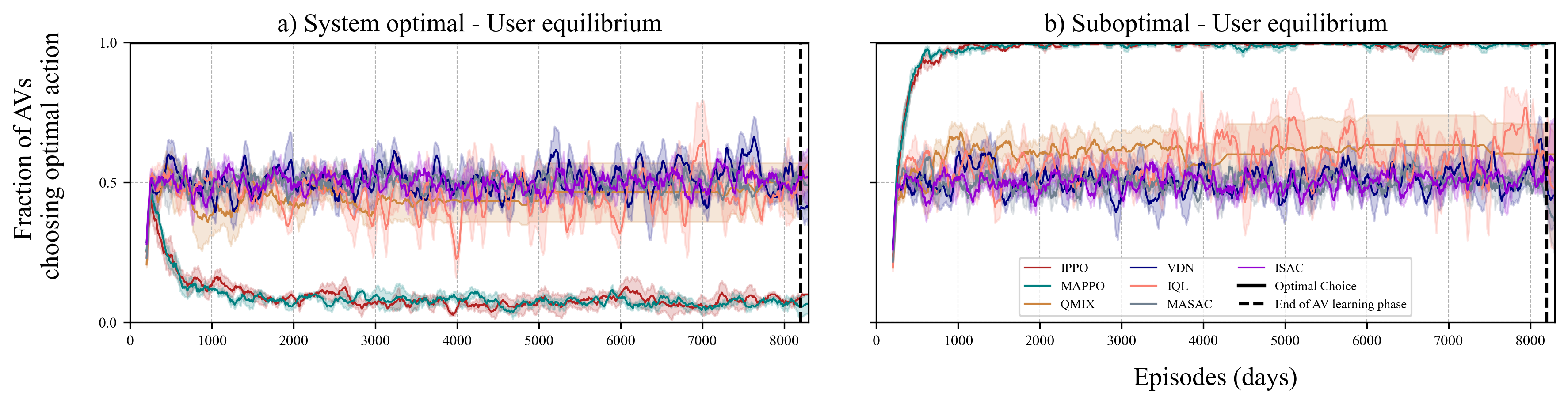}}
\caption{\textbf{8000} episodes (days) of training for selected MARL algorithms while humans adapt their choices. The trivial solution is found only by IPPO and MAPPO in the suboptimal user equilibrium. Other algorithms remain probabilistic after and fail to converge to optimal policy. The optimal solution in the suboptimal equilibrium is achieved after 1000 days. The error bars depict the 15th and 85th percentile values across replications and the solid line indicates the mean value across replications.}

\section{Results for smaller AV fleets}
\label{sec:smaller_av_fleets}

 \begin{figure}[H]
    \centering
    \begin{subfigure}[b]{0.45\textwidth}
        \centering
        \includegraphics[width=0.9\textwidth, height=0.9\textwidth, keepaspectratio,  clip]{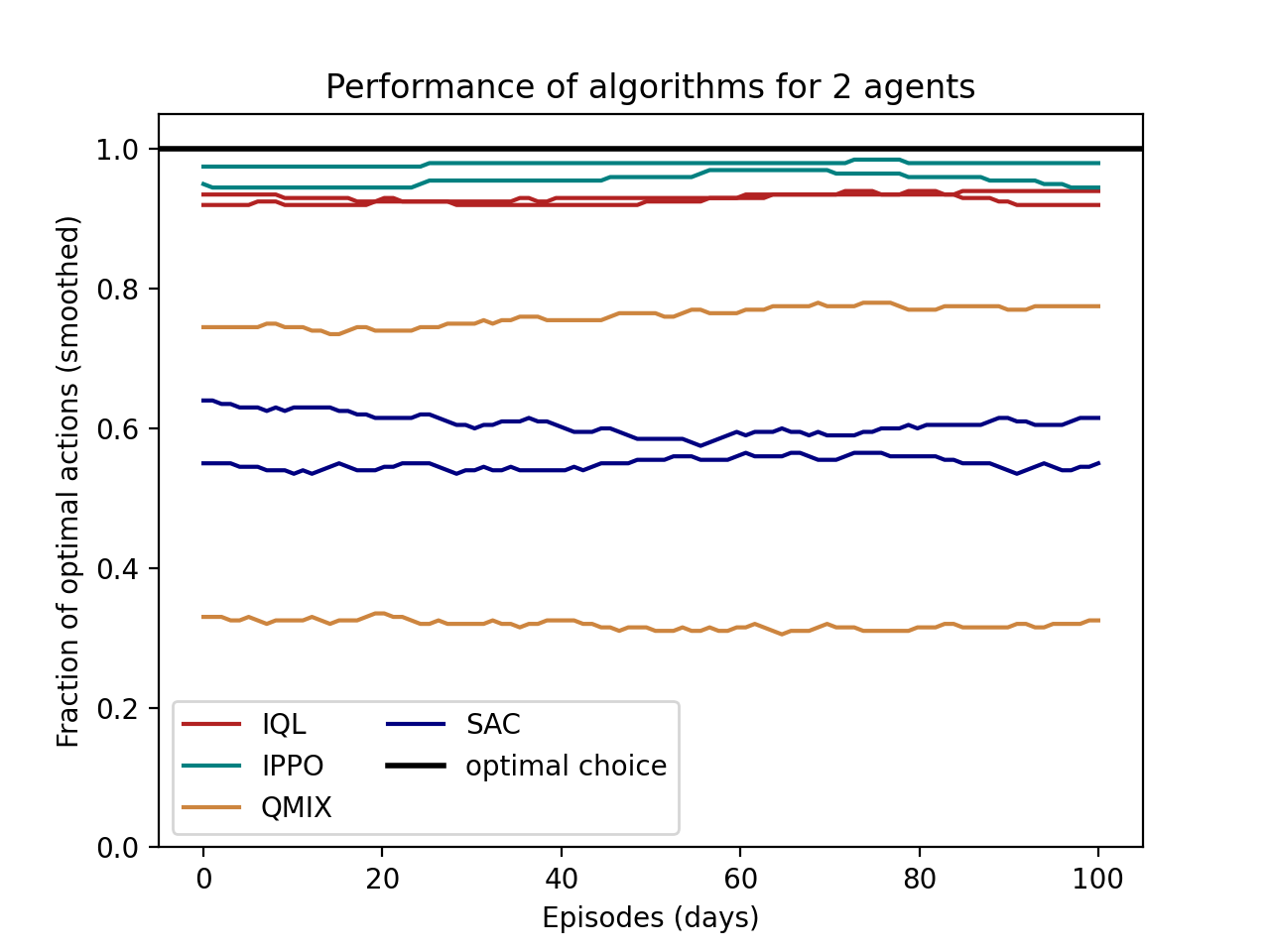}
        \caption{Performance of 2 AV fleet}
        \label{fig:2_AVs}
    \end{subfigure}
    \hfill
    \begin{subfigure}[b]{0.45\textwidth}
        \centering
        \includegraphics[width=0.9\textwidth, height=0.9\textwidth, keepaspectratio,  clip]{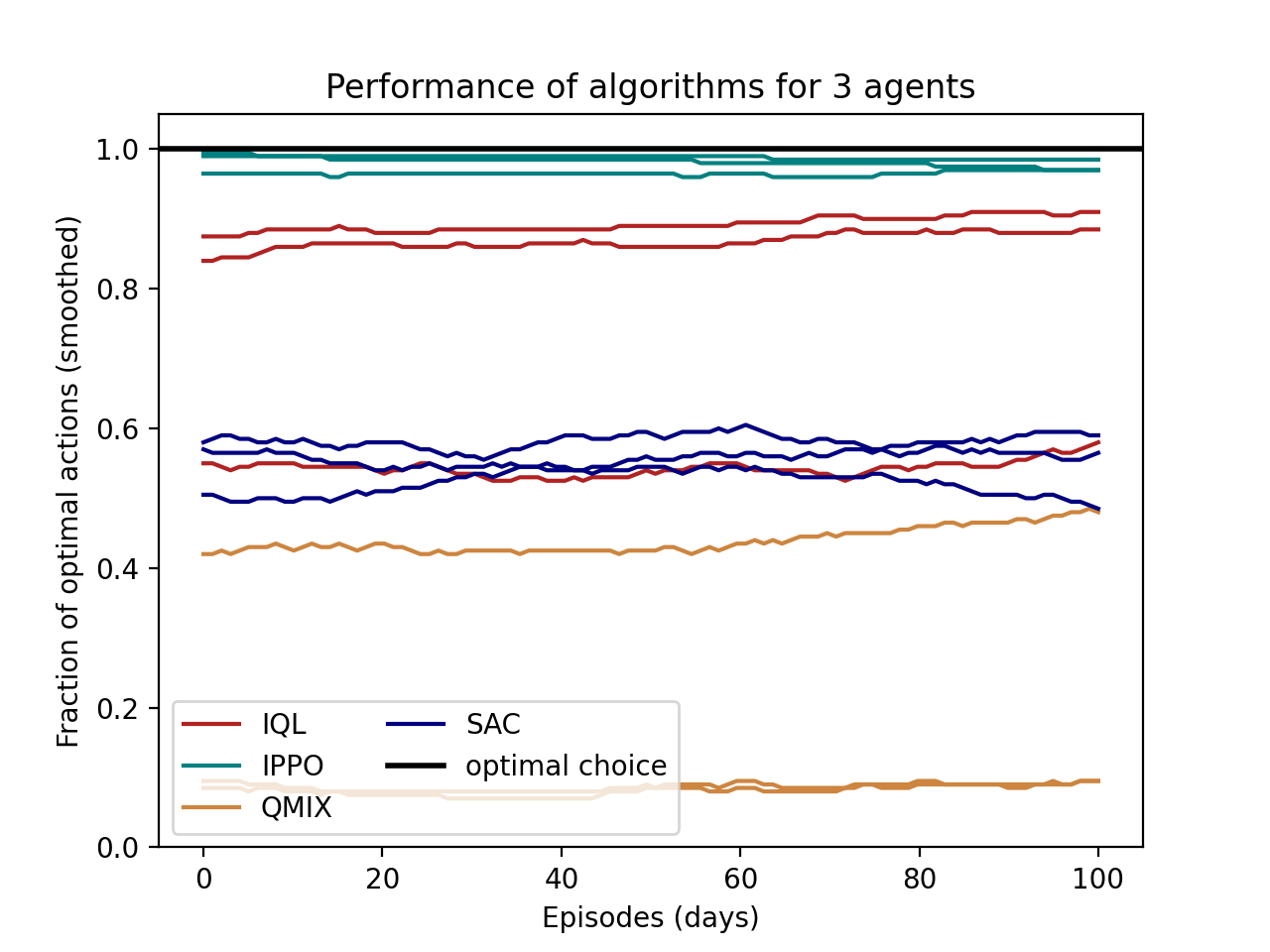}
        \caption{Performance of 3 AV fleet}
        \label{fig:3_AVs}
    \end{subfigure}

    \vspace{1em} 

    \begin{subfigure}[b]{0.45\textwidth}
        \centering
        \includegraphics[width=0.9\textwidth, height=0.9\textwidth, keepaspectratio,  clip]{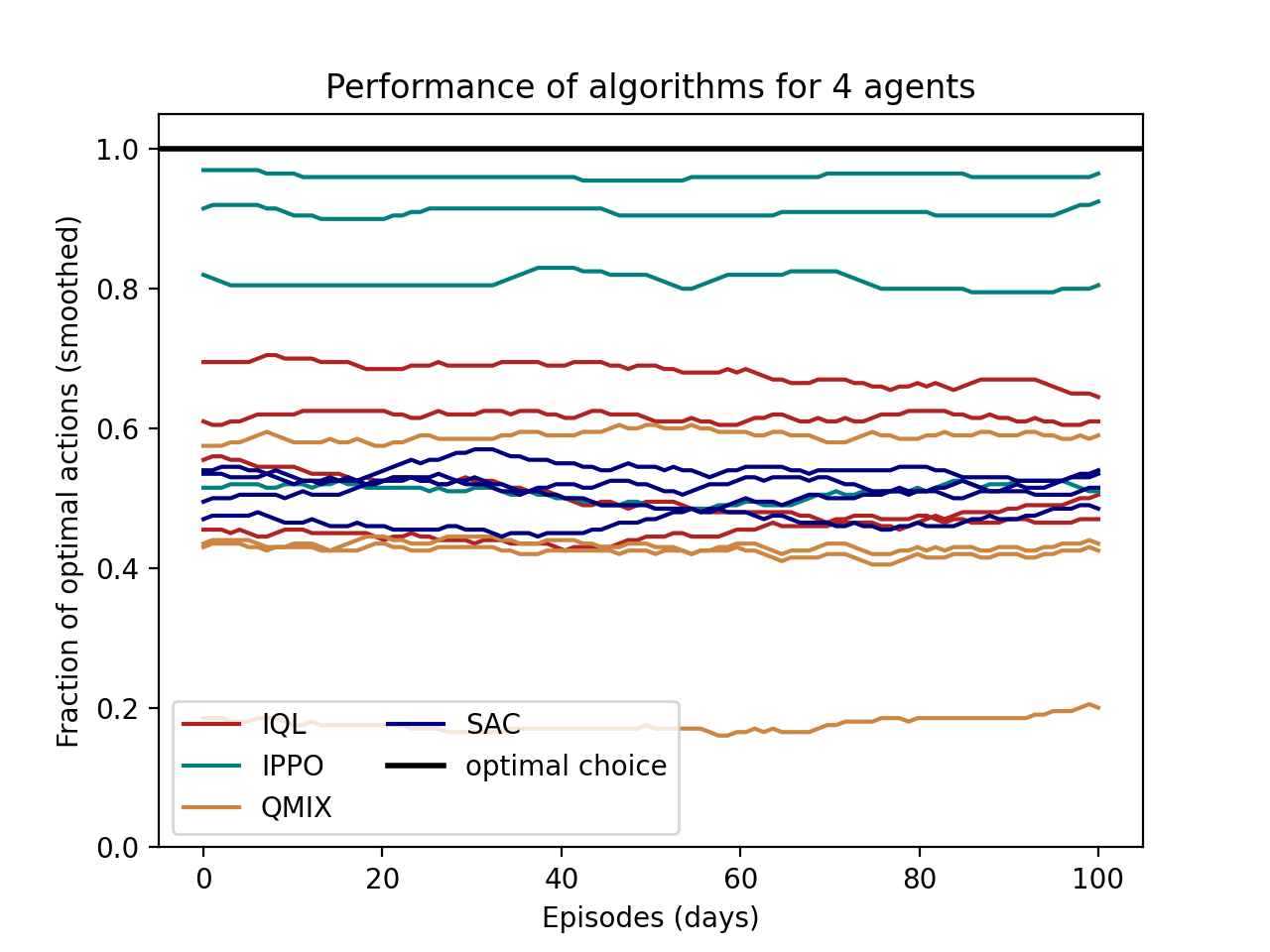}
        \caption{Performance of 4 AV fleet}
        \label{fig:4_AVs}
    \end{subfigure}
    \hfill
    \begin{subfigure}[b]{0.45\textwidth}
        \centering
        \includegraphics[width=0.9\textwidth, height=0.9\textwidth, keepaspectratio, clip]{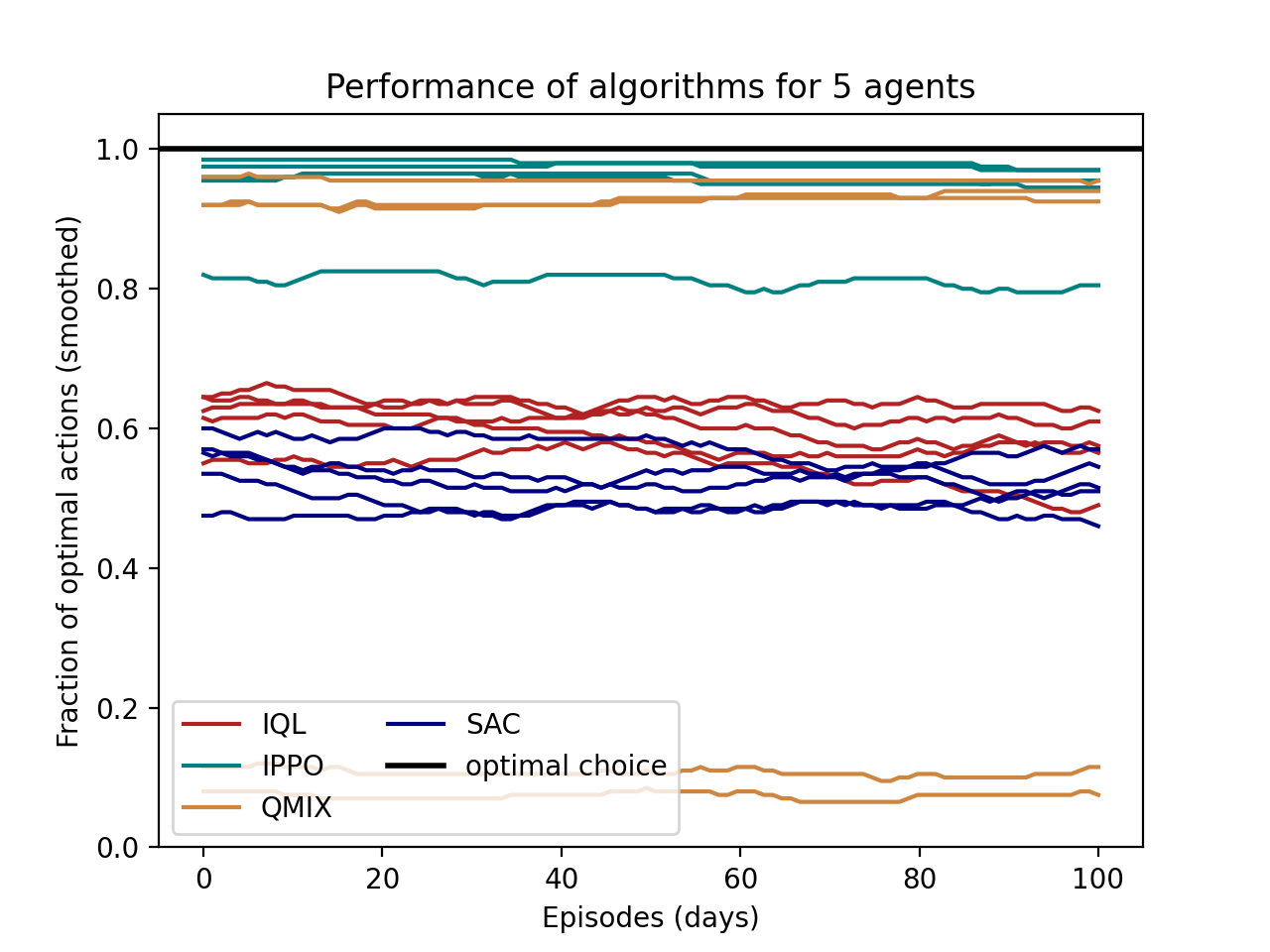}
        \caption{Performance of 5 AV fleet}
        \label{fig:5_AVs}
    \end{subfigure}

    \caption{Results (last 100 episodes - testing phase) of the randomly sampled simulations for smaller fleet sizes. Several algorithms failed to find the optimal solutions and performed worse as the number of AVs in the system increased.}
    \label{fig:appendix}
\end{figure}
\label{fig:multiple-agent-adaptation}
\end{center}
\vskip -0.1in
\end{figure}

\section{Mean rewards}

\begin{figure}[H]
    \centering
    \begin{subfigure}[b]{0.45\textwidth}
        \centering
        \includegraphics[width=\textwidth]{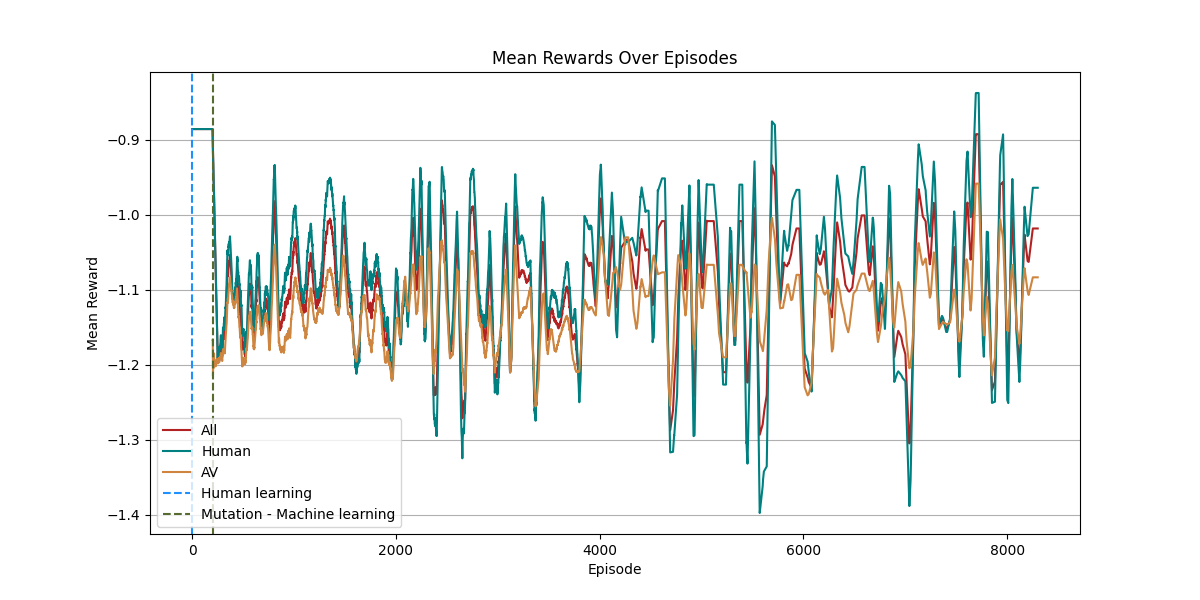}
        \caption{Independent Q-learning (IQL)}
        \label{fig:figure1}
    \end{subfigure}
    \hfill
    \begin{subfigure}[b]{0.45\textwidth}
        \centering
        \includegraphics[width=\textwidth]{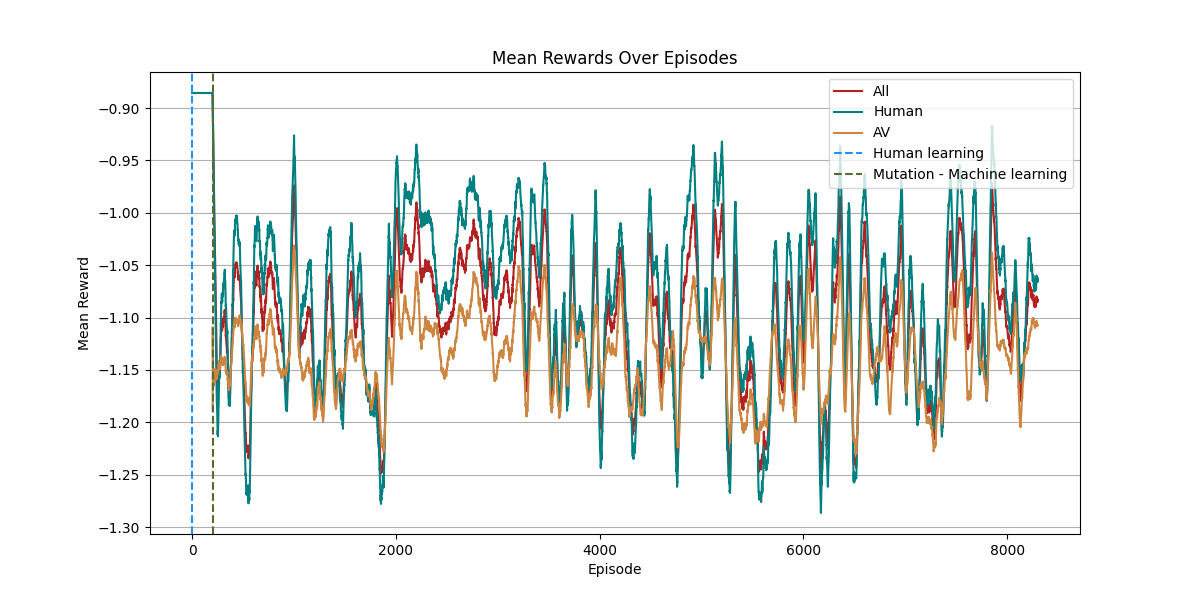}
        \caption{Value-decomposition networks (VDN)}
        \label{fig:figure2}
    \end{subfigure}

    \vspace{1em} 

    \begin{subfigure}[b]{0.45\textwidth}
        \centering
        \includegraphics[width=\textwidth]{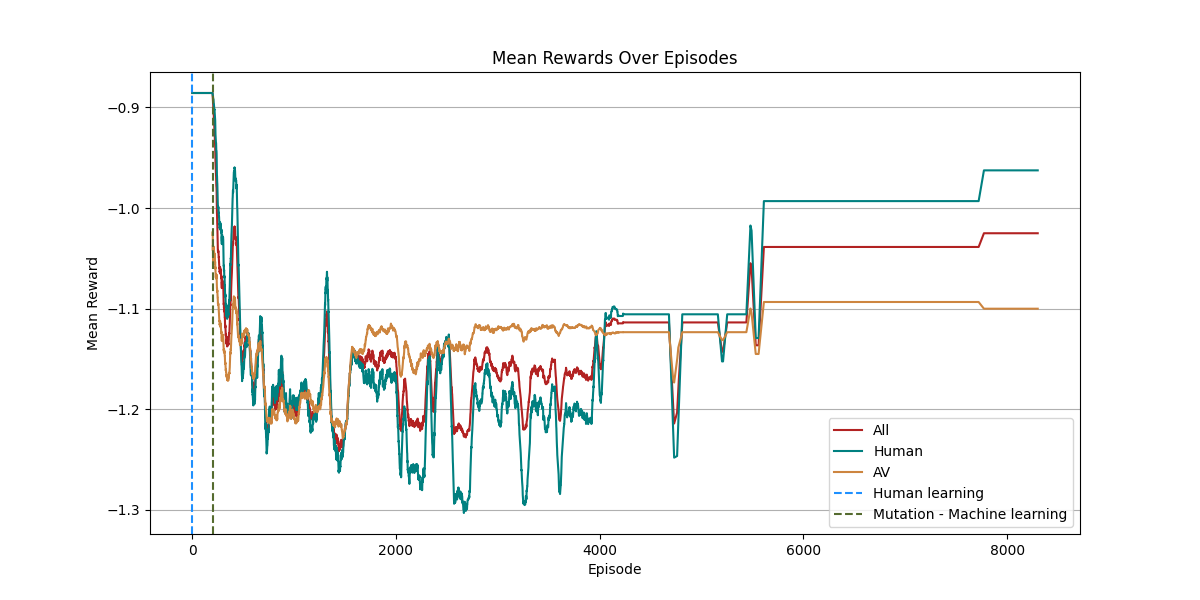}
        \caption{Monotonic value function factorisation (QMIX)}
        \label{fig:figure3}
    \end{subfigure}
    \hfill
    \begin{subfigure}[b]{0.45\textwidth}
        \centering
        \includegraphics[width=\textwidth]{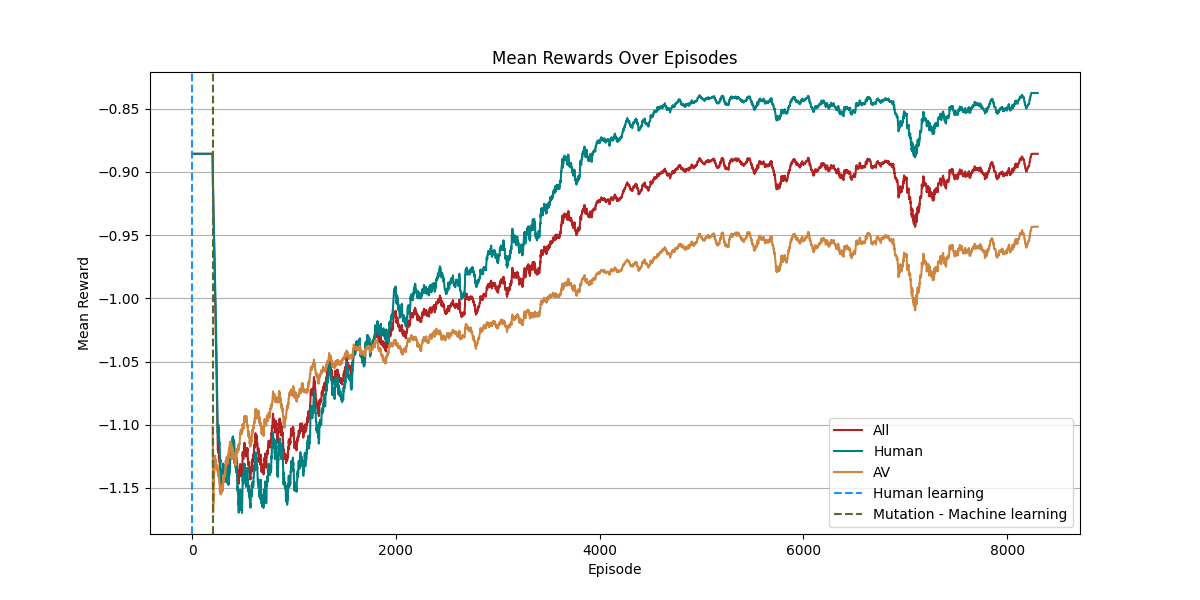}
        \caption{Multi-agent proximal policy optimization (MAPPO)}
        \label{fig:figure4}
    \end{subfigure}

    \vspace{1em} 
    
    \begin{subfigure}[b]{0.45\textwidth}
        \centering
        \includegraphics[width=\textwidth]{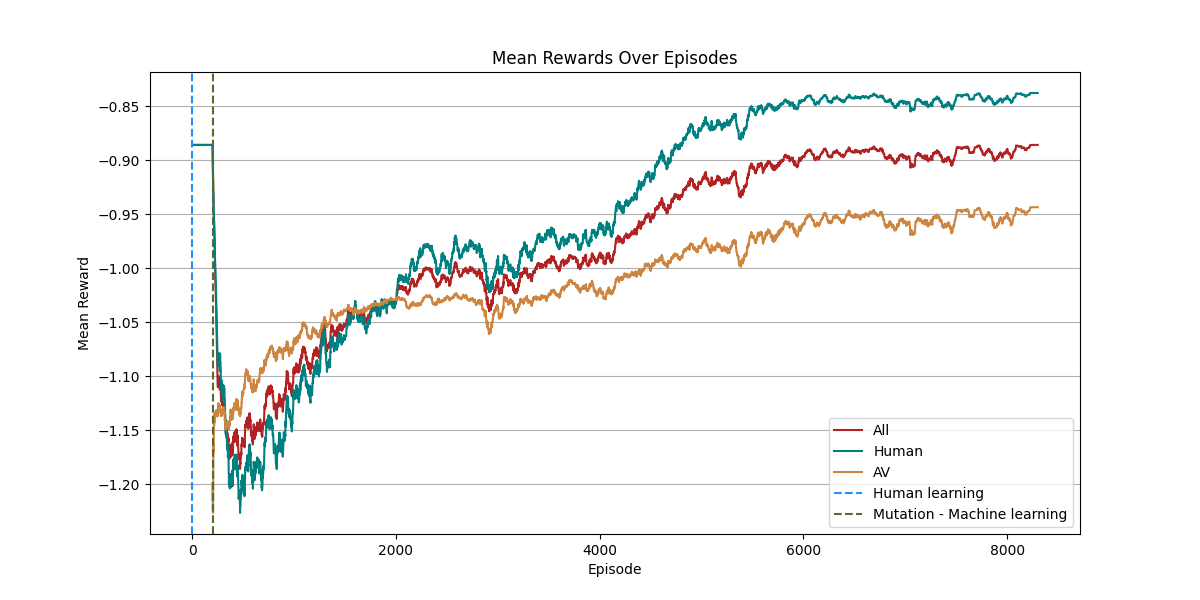}
        \caption{Independent proximal policy optimization (IPPO)}
        \label{fig:figure5}
    \end{subfigure}
    \hfill
    \begin{subfigure}[b]{0.45\textwidth}
        \centering
        \includegraphics[width=\textwidth]{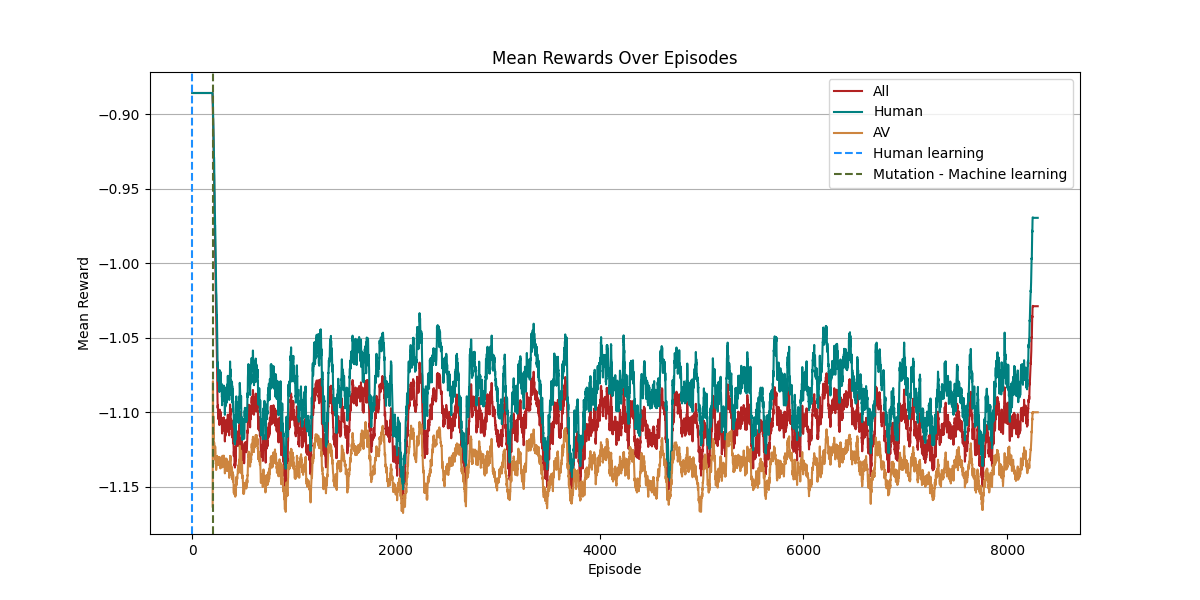}
        \caption{Independent soft actor-critic (ISAC)}
        \label{fig:figure6}
    \end{subfigure}

    \vspace{1em} 
    
    
    \begin{subfigure}[b]{0.45\textwidth}
        \centering
        \includegraphics[width=\textwidth]{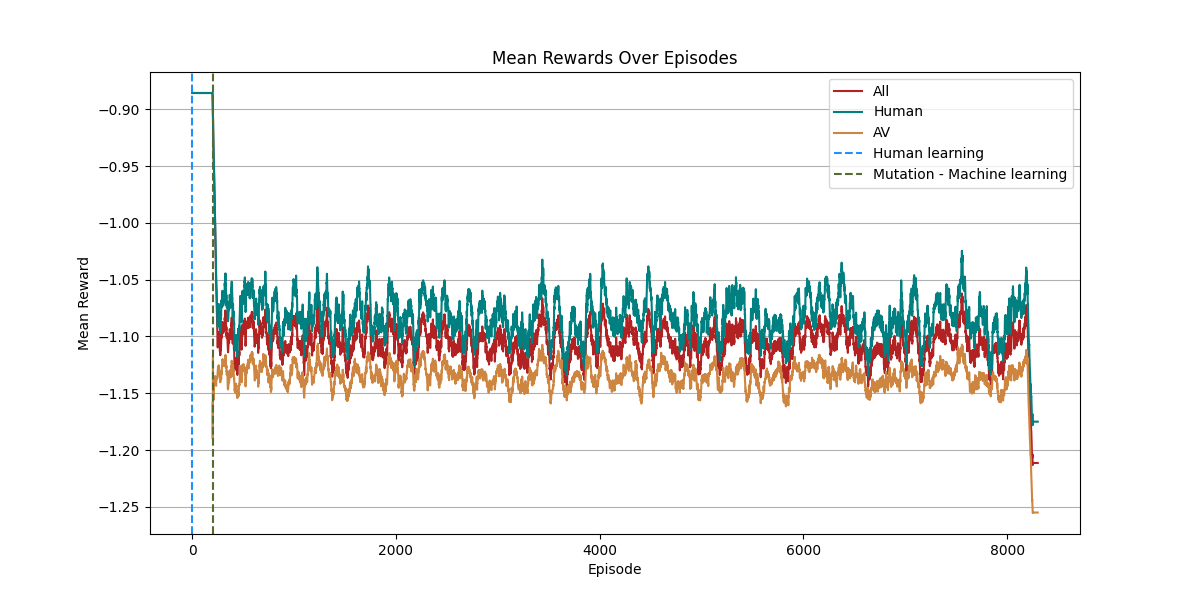}
        \caption{Multi-agent soft actor-critic (MASAC)}
        \label{fig:figure7}
    \end{subfigure}

    \caption{\textbf{Mean rewards} of AV agents, human agents, and both combined under the system optimum user equilibrium scenario. The initial 200 episodes of the simulation represent the human learning phase. The mutation event at episode 200 initiates the training of AV agents using different MARL algorithms and the end of the human learning phase. The testing phase corresponds to the last 100 episodes. Notably, IPPO and MAPPO algorithms converge after 6000 episodes to the lowest average reward, while other algorithms either stabilize at higher reward values or exhibit oscillatory behavior during training. The episodes in these plots do not correspond to policy updates (more details in Appendix Section B).}
    \label{fig:mean_rewards_zero}
\end{figure}

\begin{figure}[H]
    \centering
    \begin{subfigure}[b]{0.45\textwidth}
        \centering
        \includegraphics[width=\textwidth]{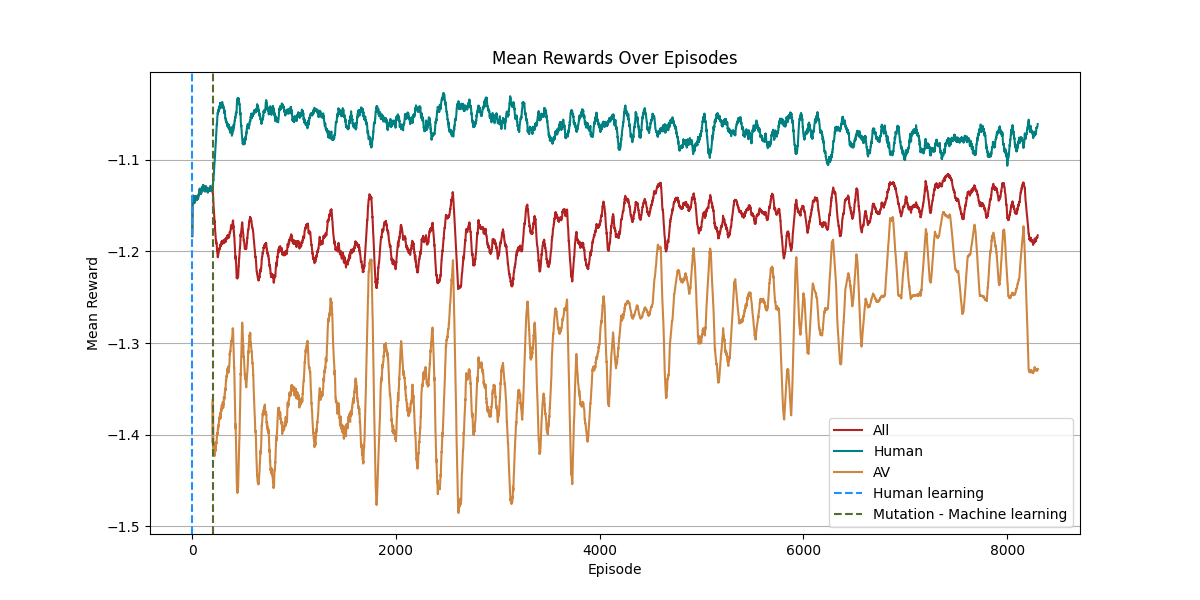}
        \caption{IQL}
        \label{fig:figure1}
    \end{subfigure}
    \hfill
    \begin{subfigure}[b]{0.45\textwidth}
        \centering
        \includegraphics[width=\textwidth]{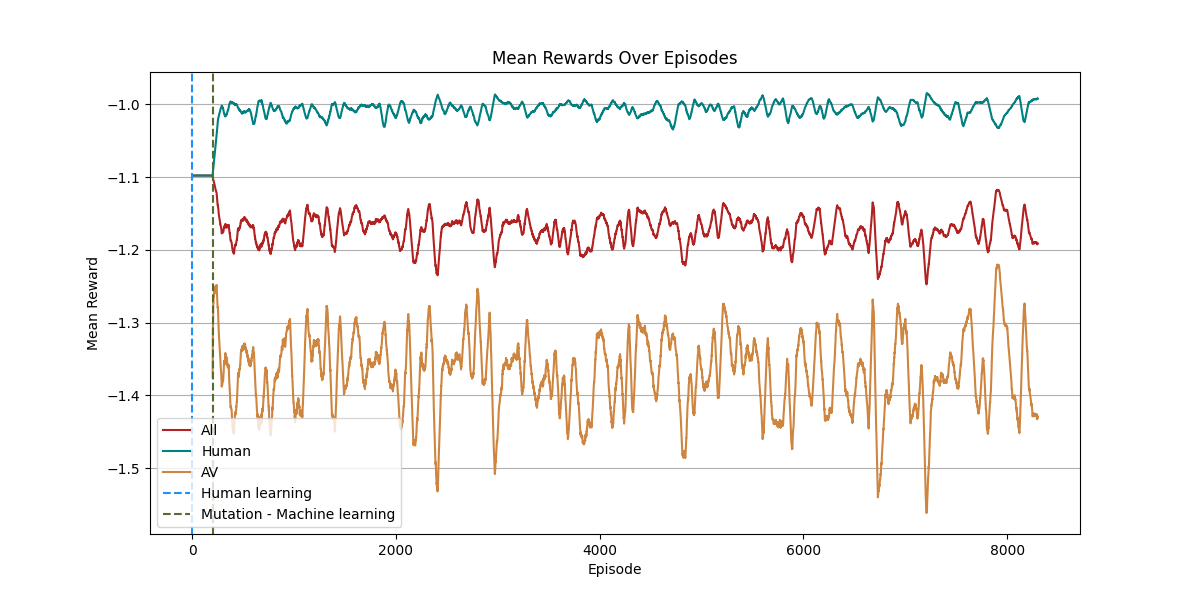}
        \caption{VDN}
        \label{fig:figure2}
    \end{subfigure}

    \vspace{1em} 

    \begin{subfigure}[b]{0.45\textwidth}
        \centering
        \includegraphics[width=\textwidth]{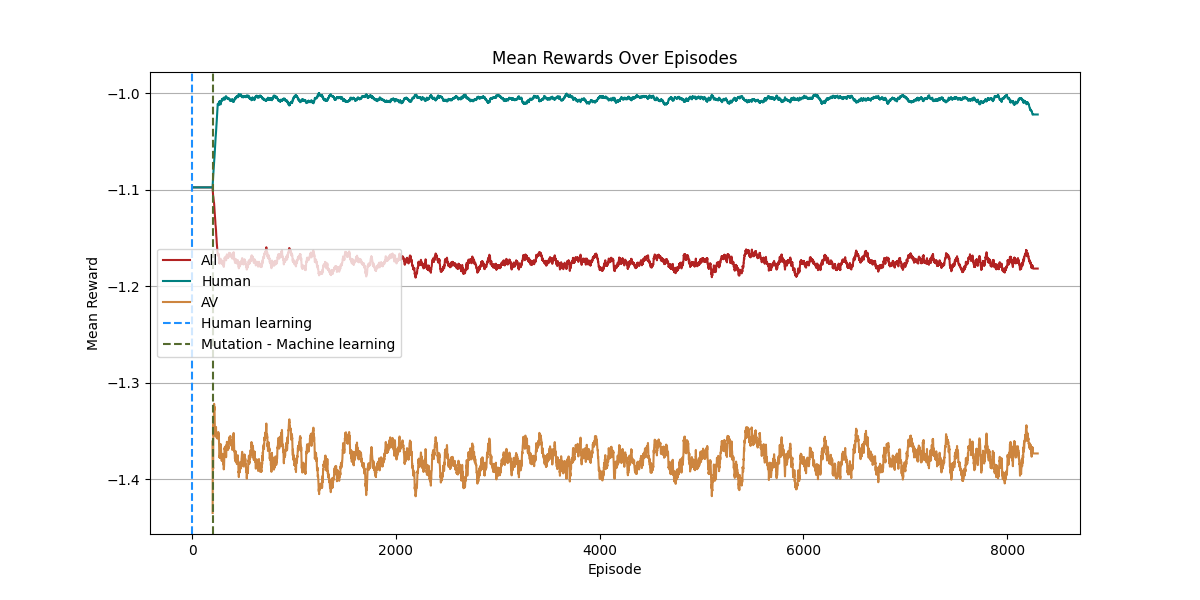}
        \caption{QMIX}
        \label{fig:figure3}
    \end{subfigure}
    \hfill
    \begin{subfigure}[b]{0.45\textwidth}
        \centering
        \includegraphics[width=\textwidth]{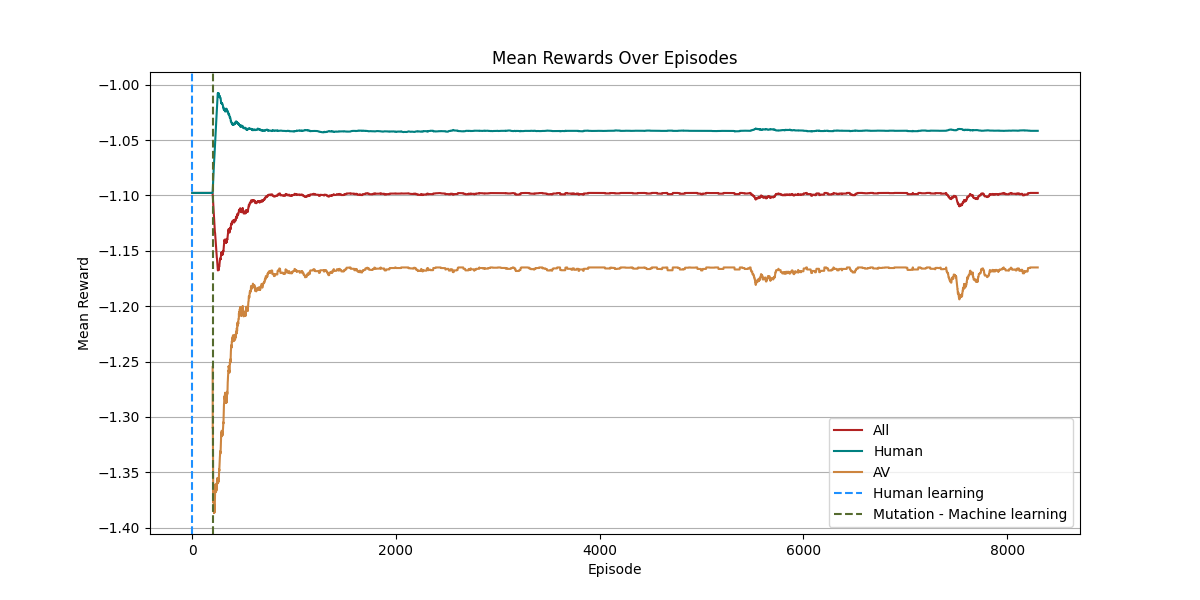}
        \caption{MAPPO}
        \label{fig:figure4}
    \end{subfigure}

    \vspace{1em} 
    
    \begin{subfigure}[b]{0.45\textwidth}
        \centering
        \includegraphics[width=\textwidth]{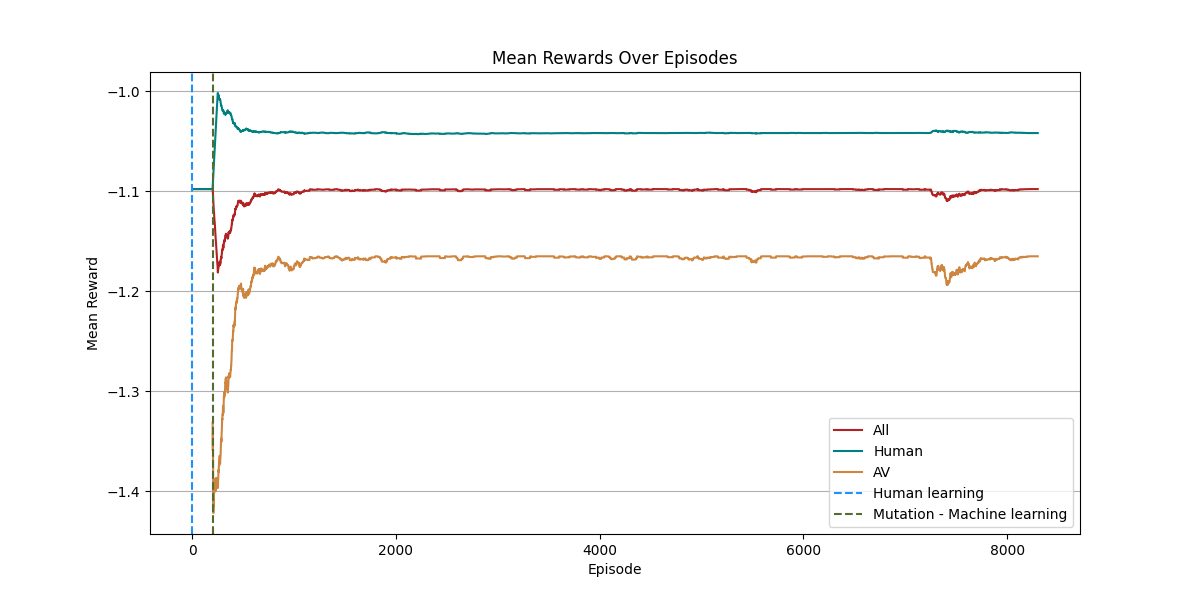}
        \caption{IPPO}
        \label{fig:figure5}
    \end{subfigure}
    \hfill
    \begin{subfigure}[b]{0.45\textwidth}
        \centering
        \includegraphics[width=\textwidth]{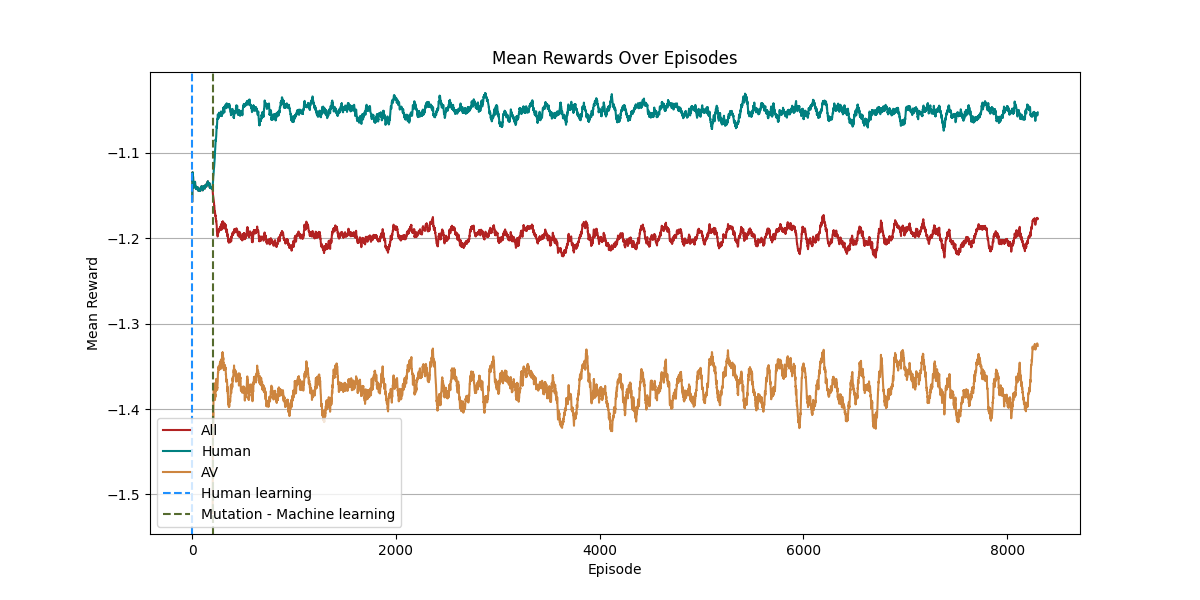}
        \caption{ISAC}
        \label{fig:figure6}
    \end{subfigure}

    \vspace{1em} 
    
    
    \begin{subfigure}[b]{0.45\textwidth}
        \centering
        \includegraphics[width=\textwidth]{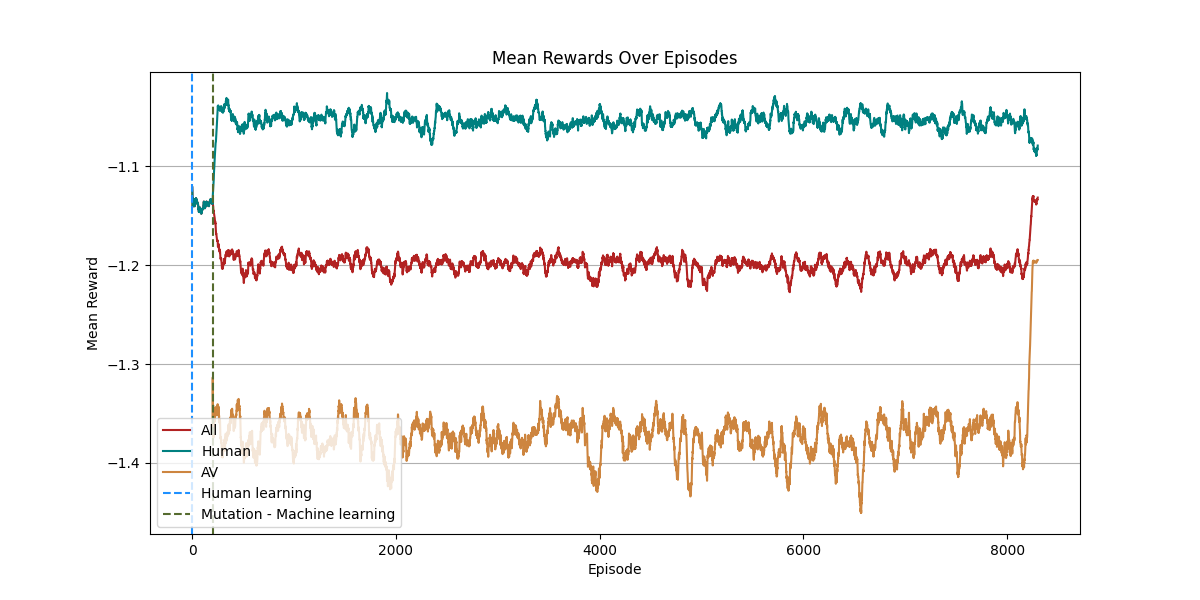}
        \caption{MASAC}
        \label{fig:figure7}
    \end{subfigure}

    \caption{\textbf{Mean rewards} of AV agents, human agents, and both combined in the suboptimal user equilibrium. IPPO and MAPPO achieve the lowest average rewards in the testing phase and during the training their reward is less oscillatory than the other algorithms. The rewards in this scenario are higher than those in \ref{fig:mean_rewards_zero}, as the system is under the suboptimal use equilibrium.}
    \label{fig:mean_rewards_one}
\end{figure}

\section{Action shifts}

\begin{figure}[H]
    \centering
    \begin{subfigure}[b]{0.45\textwidth}
        \centering
        \includegraphics[width=\textwidth, trim=0cm 0cm 15cm 0cm, clip]{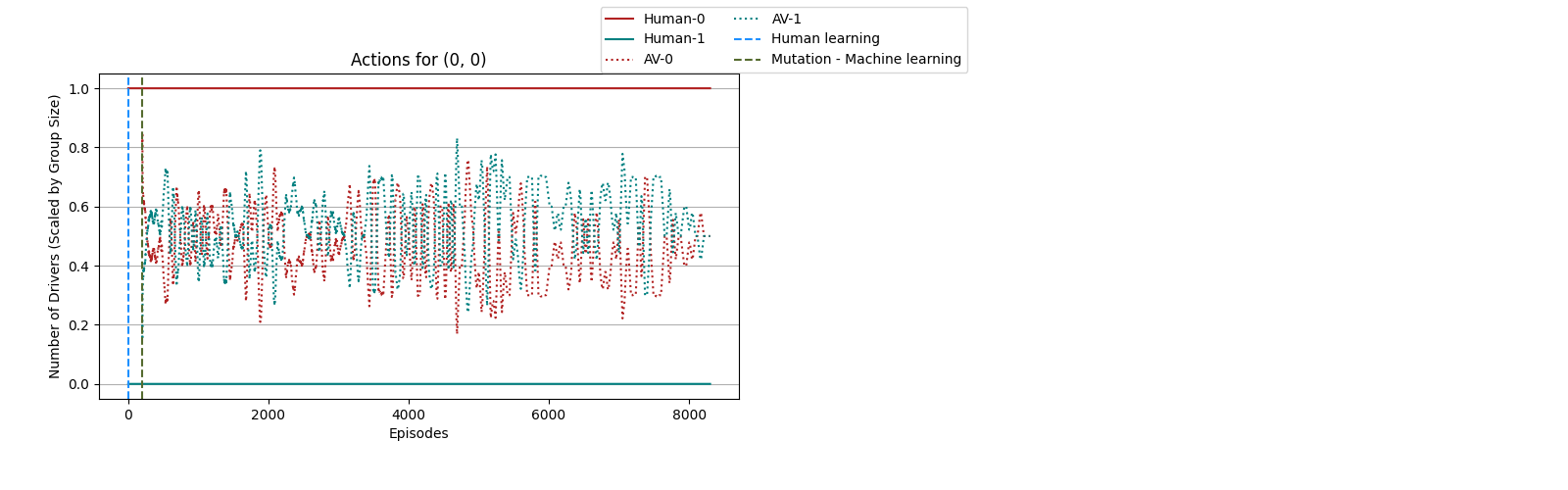}
        \caption{IQL}
        \label{fig:figure1}
    \end{subfigure}
    \hfill
    \begin{subfigure}[b]{0.45\textwidth}
        \centering
        \includegraphics[width=\textwidth, trim=0cm 0cm 15cm 0cm, clip]{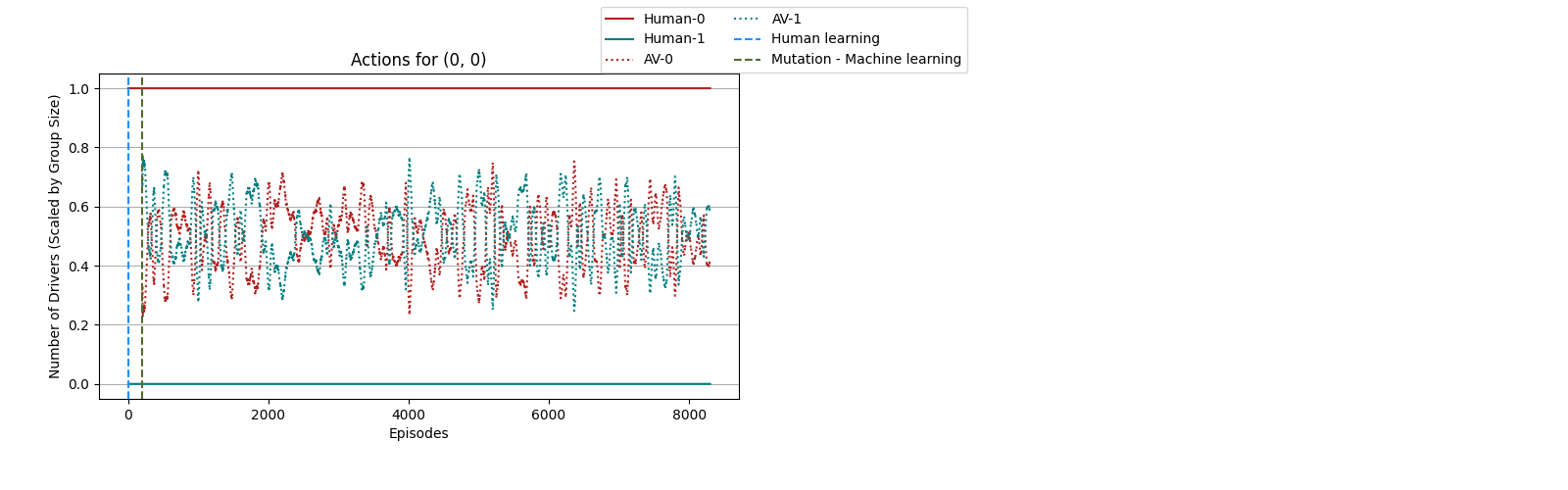}
        \caption{VDN}
        \label{fig:figure2}
    \end{subfigure}

    \vspace{1em} 

    \begin{subfigure}[b]{0.45\textwidth}
        \centering
        \includegraphics[width=\textwidth, trim=0cm 0cm 15cm 0cm, clip]{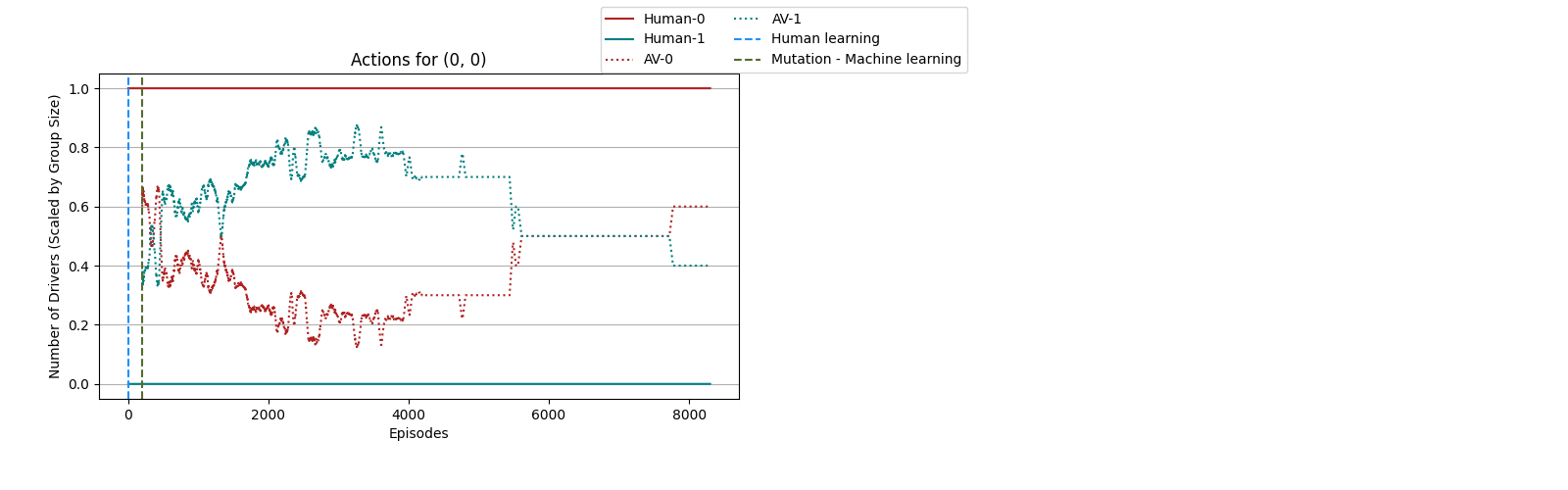}
        \caption{QMIX}
        \label{fig:figure3}
    \end{subfigure}
    \hfill
    \begin{subfigure}[b]{0.45\textwidth}
        \centering
        \includegraphics[width=\textwidth, trim=0cm 0cm 15cm 0cm, clip]{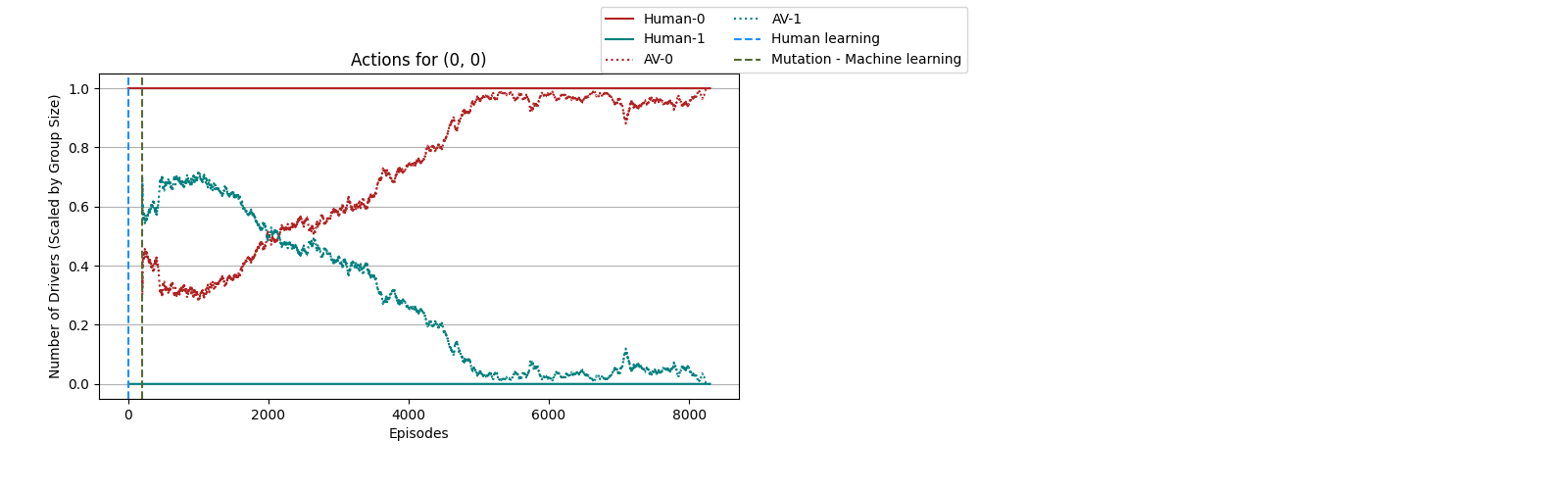}
        \caption{MAPPO}
        \label{fig:figure4}
    \end{subfigure}

    \vspace{1em} 
    
    \begin{subfigure}[b]{0.45\textwidth}
        \centering
        \includegraphics[width=\textwidth, trim=0cm 0cm 15cm 0cm, clip]{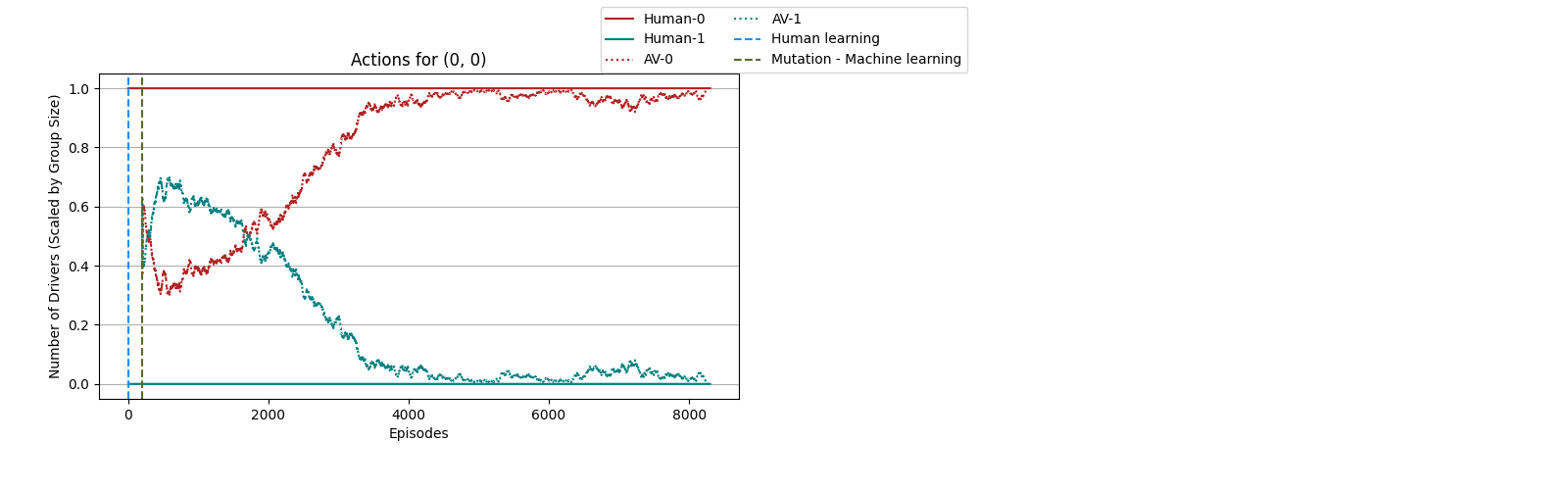}
        \caption{IPPO}
        \label{fig:figure5}
    \end{subfigure}
    \hfill
    \begin{subfigure}[b]{0.45\textwidth}
        \centering
        \includegraphics[width=\textwidth, trim=0cm 0cm 15cm 0cm, clip]{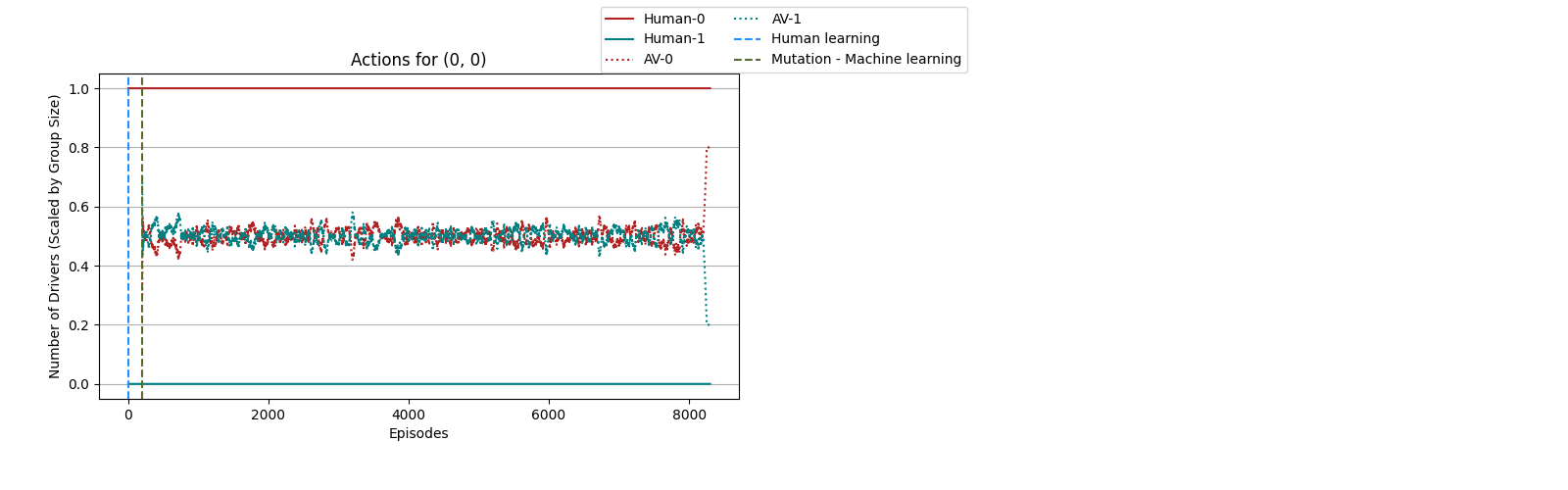}
        \caption{ISAC}
        \label{fig:figure6}
    \end{subfigure}

    \vspace{1em} 
    
    
    \begin{subfigure}[b]{0.45\textwidth}
        \centering
        \includegraphics[width=\textwidth, trim=0cm 0cm 15cm 0cm, clip]{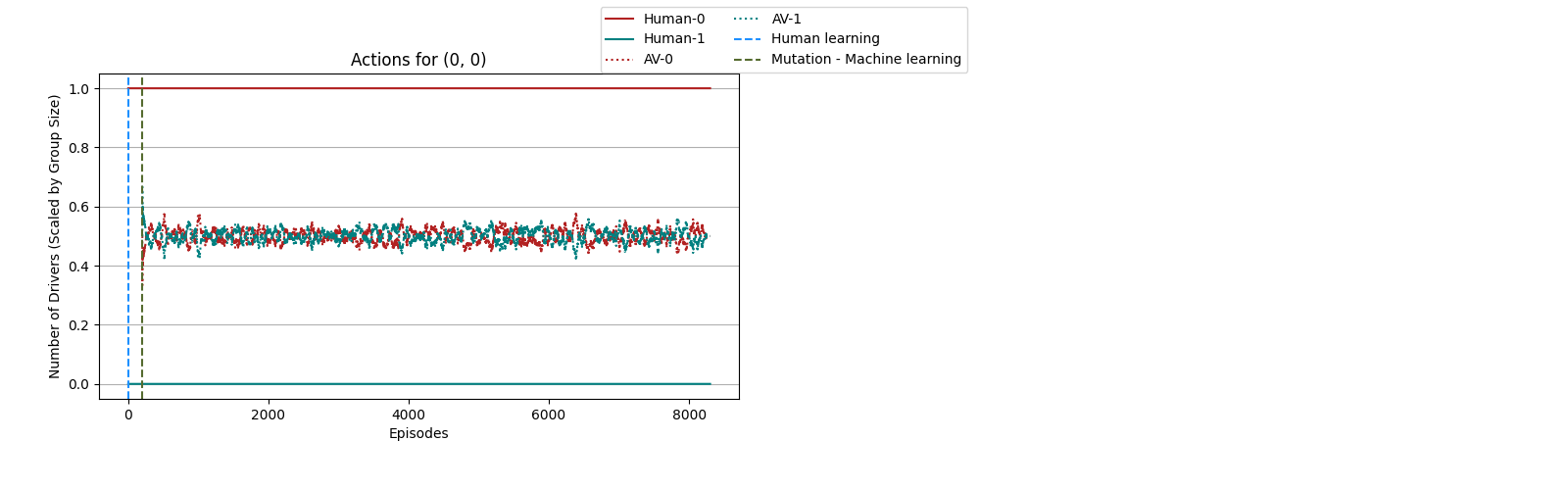}
        \caption{MASAC}
        \label{fig:figure7}
    \end{subfigure}

    \caption{\textbf{Action shifts} of AV and human agents under the system optimum user equilibrium. The plots depict the number of human agents and AVs that choose routes 1 and 0. Among the algorithms, IPPO converges the fastest to the optimal solution, where all AV agents select route 0. MAPPO also converges to this solution but requires additional episodes. The remaining algorithms settle on suboptimal solutions.}
    \label{fig:action_shifts_zero}
\end{figure}

\begin{figure}[H]
    \centering
    \begin{subfigure}[b]{0.45\textwidth}
        \centering
        \includegraphics[width=\textwidth, trim=0cm 0cm 15cm 0cm, clip]{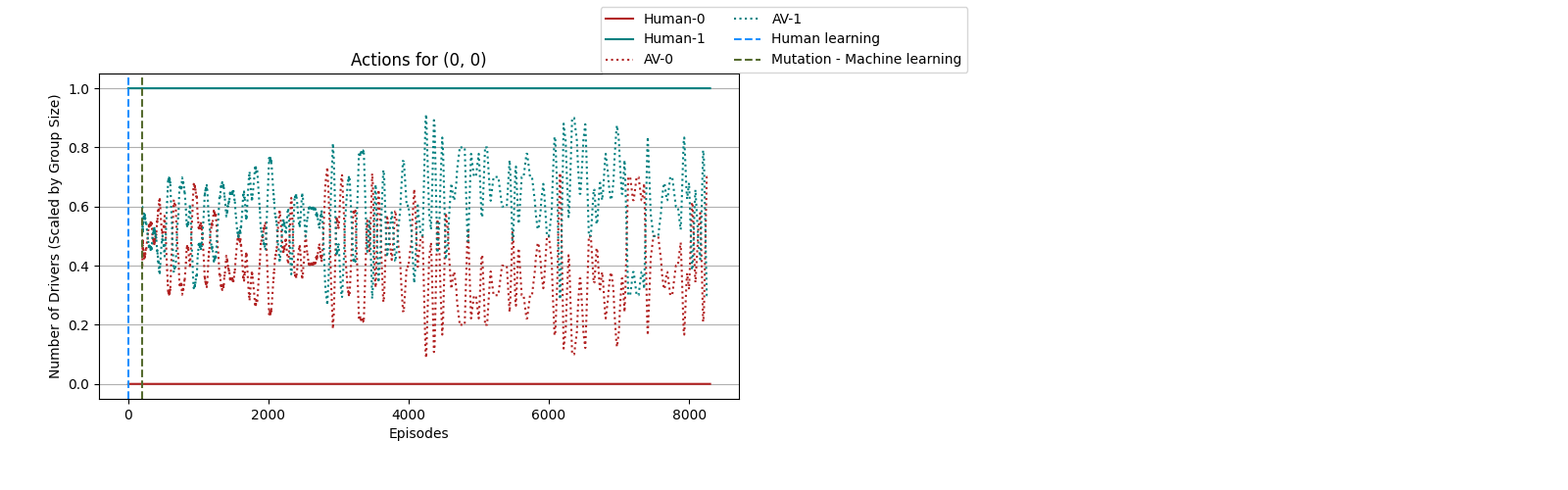}
        \caption{IQL}
        \label{fig:figure1}
    \end{subfigure}
    \hfill
    \begin{subfigure}[b]{0.45\textwidth}
        \centering
        \includegraphics[width=\textwidth, trim=0cm 0cm 15cm 0cm, clip]{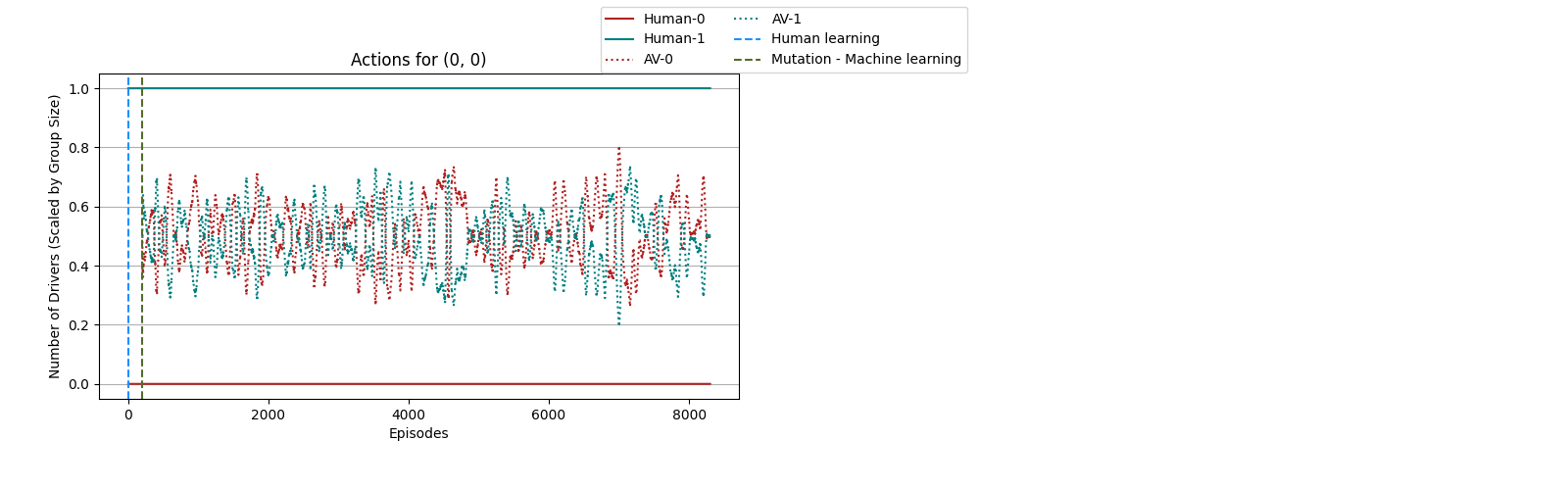}
        \caption{VDN}
        \label{fig:figure2}
    \end{subfigure}

    \vspace{1em} 

    \begin{subfigure}[b]{0.45\textwidth}
        \centering
        \includegraphics[width=\textwidth, trim=0cm 0cm 15cm 0cm, clip]{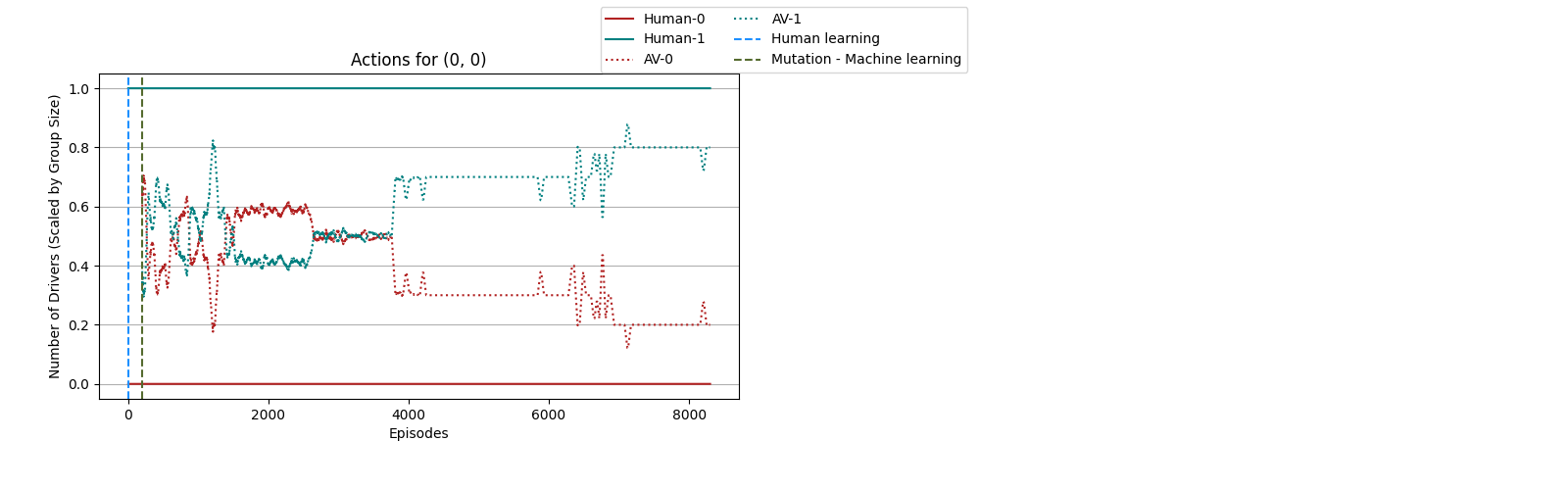}
        \caption{QMIX}
        \label{fig:figure3}
    \end{subfigure}
    \hfill
    \begin{subfigure}[b]{0.45\textwidth}
        \centering
        \includegraphics[width=\textwidth, trim=0cm 0cm 15cm 0cm, clip]{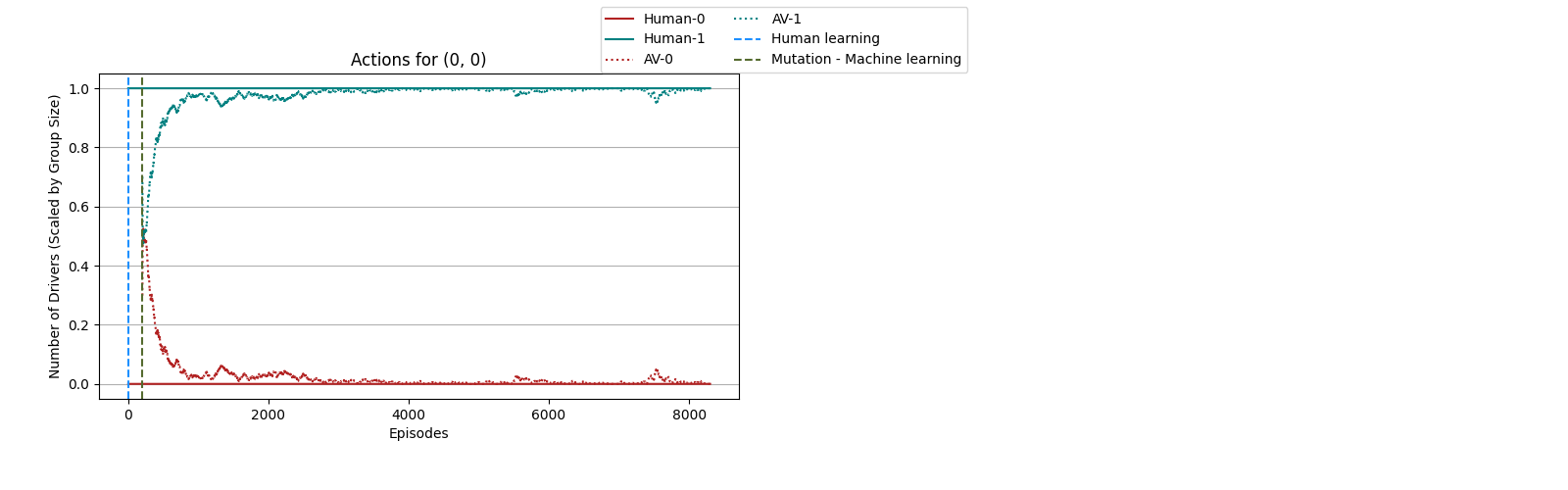}
        \caption{MAPPO}
        \label{fig:figure4}
    \end{subfigure}

    \vspace{1em} 
    
    \begin{subfigure}[b]{0.45\textwidth}
        \centering
        \includegraphics[width=\textwidth, trim=0cm 0cm 15cm 0cm, clip]{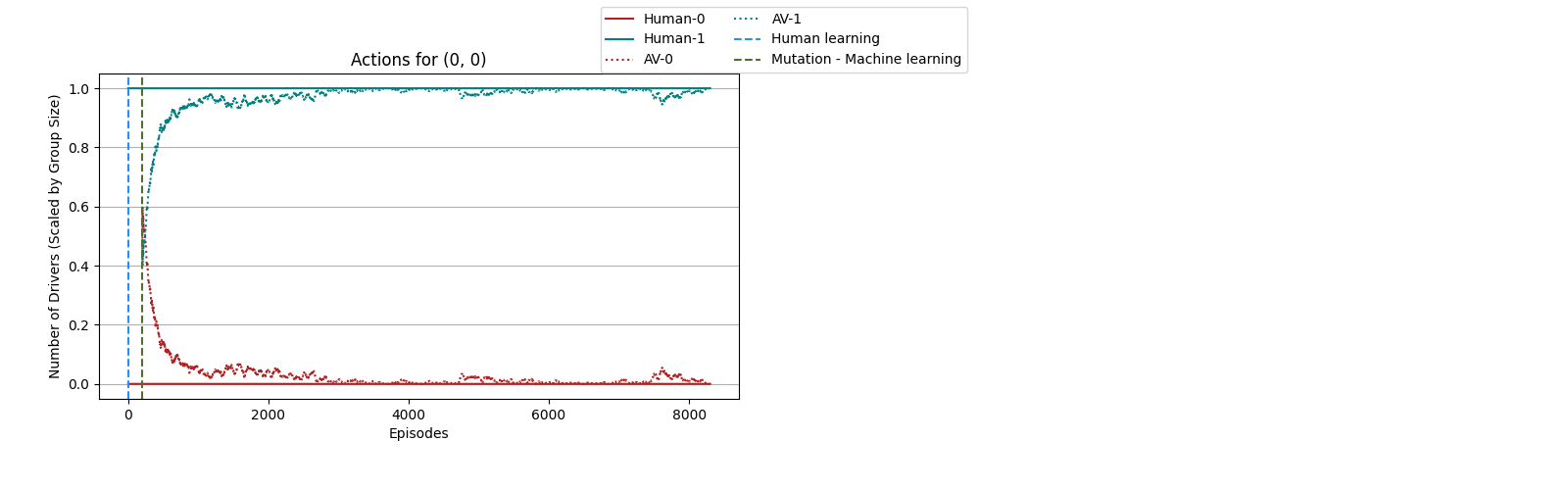}
        \caption{IPPO}
        \label{fig:figure5}
    \end{subfigure}
    \hfill
    \begin{subfigure}[b]{0.45\textwidth}
        \centering
        \includegraphics[width=\textwidth, trim=0cm 0cm 15cm 0cm, clip]{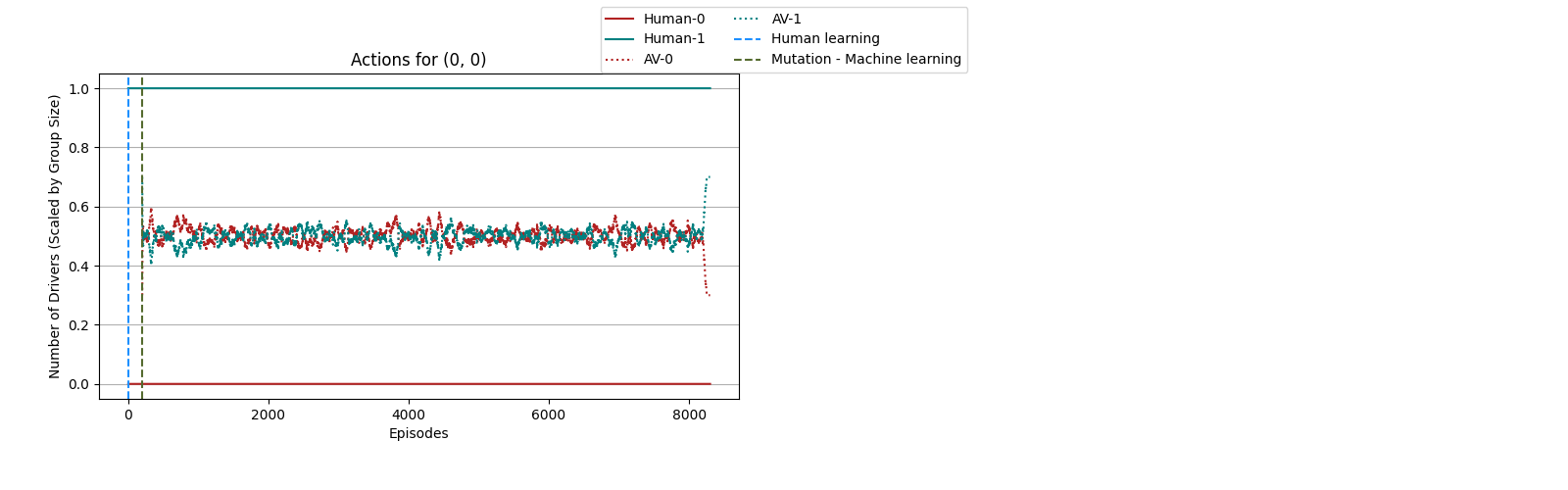}
        \caption{ISAC}
        \label{fig:figure6}
    \end{subfigure}

    \vspace{1em} 
    
    
    \begin{subfigure}[b]{0.45\textwidth}
        \centering
        \includegraphics[width=\textwidth, trim=0cm 0cm 15cm 0cm, clip]{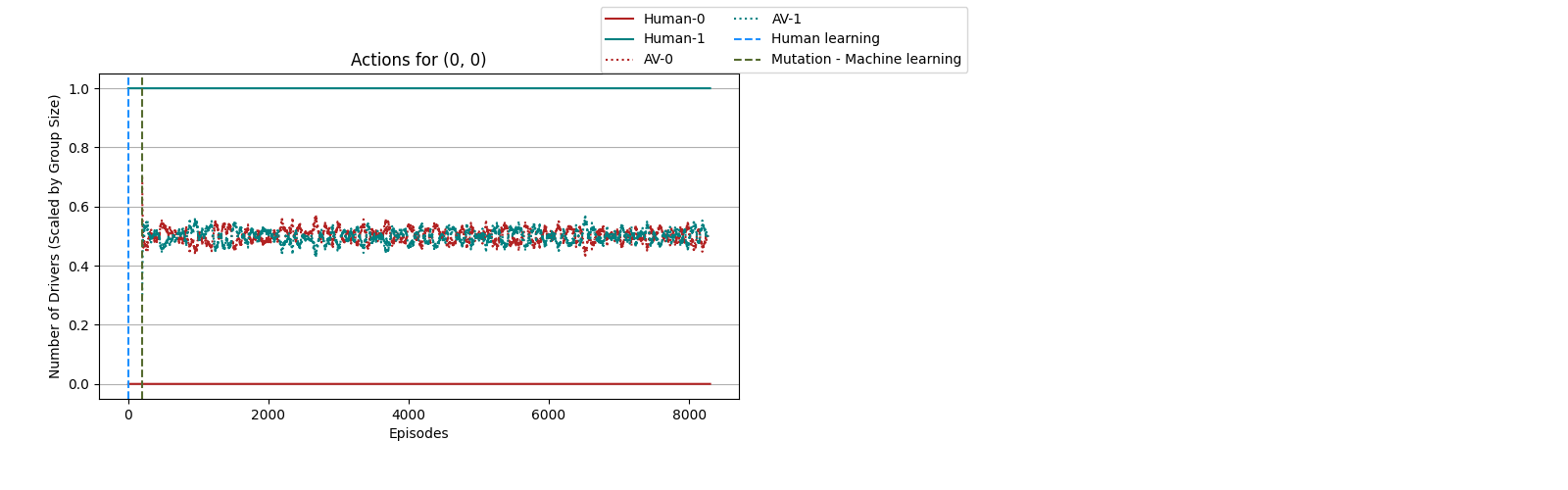}
        \caption{MASAC}
        \label{fig:figure7}
    \end{subfigure}

    \caption{\textbf{Action shifts} of AV and human agents in the suboptimal user equilibrium. The plots illustrate the number of human agents and AVs selecting routes 0 and 1. Among algorithms, IPPO and MAPPO converge even faster to the optimal solution, compared to the system optimum user equilibrium scenario, as this equilibrium is more stable. The rest of the algorithms converge to suboptimal solutions.}
    \label{fig:action_shifts_one}
\end{figure}

\end{document}